\documentclass[11pt]{ar}
\usepackage[english]{babel}
\usepackage{psfig,graphicx}
\usepackage{supertab}

\onecolumn
\begin{document}

\title{The evolutionary stage of the semiregular variable  QY\,Sge\,=\,IRAS\,20056+1834}

\author{V.G.~Klochkova,  V.E.~Panchuk, E.L. Chentsov, M.V.~Yushkin}

\date{\today}	     

\institute{Special Astrophysical Observatory RAS, Nizhnij Arkhyz,  369167 Russia}

\abstract{Repeated spectral observations made with the 6-m telescope of
SAO RAS yielded new data on the radial velocity variability of the
peculiar yellow supergiant QY\,Sge. The strongest and most peculiar
feature in its spectrum is the complex profile of Na{\sc i} D lines, which
contains a narrow and a very wide emissions. The wide emission can be seen
to extend from $-170$ to $+120$\,km/s, and at its central part it is cut
by an absorption feature, which, in turn, is split into two subcomponents
by a narrow (16\,km/s at $r$=2.5) emission. An analysis of all the
$V_r$ values leads us to adopt for the star a systemic velocity of
$V_r$=$-21.1$\,km/s, which corresponds to the position of the narrow
emission component of Na{\sc i}. The locations of emission features
of Na{\sc i} D lines are invariable, which point to their formation in
regions that are external to the supergiant's photosphere. Differential
line shifts of about 10\,km/s are revealed. Emission in the H$\alpha$ line
is weaker than in Na{\sc i} D lines, it fills the photospheric absorption
almost completely. The absorption lines in the spectrum of QY\,Sge have a
substantial width of $FWHM\approx45$\,km/s. The method of model
atmospheres is used to determine the following parameters: the effective
temperature $T_{eff}$=6250$\pm150$\,K, surface gravity $lg\,g$=2.0$\pm0.2$,
and microturbulence velocity $\xi_t$=4.5$\pm0.5$\,km/s. The chemical
composition of the atmosphere differs only slightly from the solar
composition: the metallicity of the star is found to be somewhat higher
than the solar metallicity with an average overabundance of iron-peak
elements of $[Met/H]_{\odot}$=+0.20. The star is found to be slightly
overabundant in carbon and nitrogen, $[C/Fe]$=+0.25, $[N/Fe]$=+0.27. The
$\alpha$-process elements Mg, Si, and Ca are slightly overabundant, on the
average by $[\alpha/H]_{\odot}$=+0.12, and sulfur overabundance is higher,
$[S/\alpha]$=+0.29. The strong overabundance of sodium, $[Na/Fe]$=+0.75,
is likely to be due to the dredge-up of the matter processed in the NeNa
cycle. Heavy elements of the $s$-process are underabundant relative to the
Sun. On the whole, the observed properties of QY\,Sge do not give grounds
for including this star into the group of R\,CrB or RV\,Tau-type objects. }

\authorrunning{Klochkova et al.}
\titlerunning{Evolutionary stage of QY\,Sge}

\maketitle

\section{Introduction}

The General Catalog of Variable Stars [\cite{GCVS}] classifies the yellow
supergiant QY\,Sge (Sp=G0e) as a semiregular variable. The QY\,Sge star,
associated with the IR source IRAS\,20056+1834, is located outside the
Galactic plane: $b$=$-7\lefteqn{.}^{\rm o}46$. $J$- and $H$-band polarimetric
observations [\cite{Gledhill}] that were performed with UKIRT telescope
revealed no extended structure or deviations from symmetry at the angular
resolution of FWHM=0.67$^{''}$. Polarization is constant throughout the
entire image of the object, but decreases substantially with wavelength:
$P_J$=14\%, $P_H$=7\%. Gledhill et al. [\cite{Gledhill}] classified
IRAS\,20056+1834 as a protoplanetary nebula (PPN) with a well-defined core
(a ``core-dominated object'') and a compact optically thick envelope. We so
far lack a conclusive answer to the question about the physical
mechanism of polarization in QY\,Sge in the absence of an extended
envelope, however, submillimeter-wave observations with the 15-m James
Clarke Maxwell Telescope (JCMT) also confirm the lack of such an envelope
[\cite{Gledh}]. This led Gledhill et al. [\cite{Gledh}] to conclude that
QY\,Sge is at the initial stage of the formation of its envelope. This
conclusion is consistent with the high temperature ($T_e \approx 600$\,K)
of the interstellar dust [\cite{Menz}].

The attention that spectroscopists devote to the supergiant QY\,Sge is due
to its complex optical spectrum, which includes photospheric and
interstellar components, and, first and foremost, to the strong wide
emission components of the D lines of Na{\sc i} found even in the
low-resolution spectra of the star [\cite{Menz}]. Kameswara~Rao et al.
[\cite{Rao}] were the first to study a high-resolution optical spectrum of
QY\,Sge. The above authors identified the main spectral components (the
absorption spectrum of the G-type supergiant combined with narrow
low-excitation emission features and wide emission Na{\sc i} D lines) and
determined the chemical abundances of the atmosphere of the supergiant.
They proposed a model of the system with a circumstellar torus and bipolar
mass outflow. According to this model, the central star is completely
obscured from the observer and we see the radiation reflected from the
inner wall of the torus.

There is currently no consensus of opinion concerning the status of
QY\,Sge (the proposed versions are: a massive supergiant, RV\,Tau or
R\,CrB-type star, or a spectroscopic binary), its distance (the available
estimates differ by one order of magnitude), and understanding of all the
peculiarities of its optical spectrum. That is why we decided to
spectroscopically monitor QY\,Sge with the 6-m telescope of the Special
Astrophysical Observatory of the Russian Academy of Sciences (SAO RAS) and
report the results of the first six years of observations in this paper.

\section{Observations and data reduction}

QY\,Sge is a rather faint object ($V$=$12\lefteqn{.}^{\rm m}37$) for high
resolution spectroscopy. We taken spectroscopic data with the PFES and NES
CCD--echelle spectrographs of the 6\,m telescope of SAO RAS. The
spectrograph PFES [\cite{pfes}] is mounted in the prime focus of the
telescope and we used this spectrograph in 1998 to take the spectrum of
QY\,Sge with a resolution of $R$=$17000$ and a rather high signal-to-noise
ratio ($S/N \approx$\,100 ). We used the NES spectrograph in 1998, 2002,
2003, and 2004 to take spectra with a resolution of $R$=$60000$, but with
limited S/N (S/N$\approx$40--50). See [\cite{nes1, nes2,nes3,nes4}] for
the description of the successive stages of the development and
improvement of the NES spectrograph. The spectrograph  NES uses an image
slicer at its entry to cut the image into three slices to provide a more
than double gain in flux [\cite{nes3}]. Local corrector [\cite{nes5}]
ensures the accurate centering of the star's image relative to these
slices. Wavelength calibration is performed using the central component
of the echelle orders of the spectrum of the ThAr lamp and standard
procedures of the {\sc echelle} context of {\sc midas} system for the
reduction of CCD images.

The upper rows of Table\,\ref{RV} give the observing dates and operating
wavelength region. The data obtained by Kameswara~Rao et al. [\cite{Rao}]
complement fairly well the intensity and radial-velocity measurements of
spectral lines given in the next rows of the table: the above authors took
their spectra during the summer seasons of 1999, 2000, and 2001, which we
missed, and have the same spectral resolution ($R$=60000) as our
observations.

We developed a dedicated code package [\cite{reduct}] based on the 
ECHELLE context to modify the reduction of two-dimensional spectra taken
with echelle spectrographs equipped with an image slicer. We extract each
portion of a spectral order separately and then coadd the extracted
components of the spectral order with the allowance for the wavelength
shifts due to the design features of the image slicer. We determine the
shift of each satellite slice of echelle orders along the dispersion
of the echelle by cross-correlating with the emission spectrum of the
hollow-cathode ThAr lamp also taken with an image slicer. Note that the
shifts of the slices of the echelle orders along the dispersion of the
echelle allows more efficient removal of both cosmic-ray hits and CCD
defects (``hot'' pixels, ``traps'', zero-sensitivity columns), because the
same spectral features are located on different pixels of the detectors in
the different components of spectral orders.

We use telluric lines to correct the dispersion curves. The residual
systematic errors in radial velocities do not exceed 2\,km/s for the
spectra taken with PFES, and are lower than 1\,km/s for other spectra. We
estimated the measurement errors in pure form from the spectra taken with
a one-day interval, on July 27 and 28, 2002. The radial velocities $V_r$
measured at these observing times do not differ systematically and the
mean errors range from 0.5\,km/s for the strongest (0.4$< r <$ 0.55) to
1.0\,km/s for the weakest (0.85$ < r <$0.8) lines.

\section{Peculiarities of the optical spectrum and radial velocities}

Our observations confirmed the presence of main features earlier
found in the optical spectrum of  QY\,Sge: all our spectra show a
very narrow, $\Delta\lambda=16$\,km/s, and a wide, $\Delta\lambda
\approx$\,290\,km/s, emission components of Na{\sc i} D lines.
Metal absorptions are very wide: they are two-to-three times wider
than the narrow Na{\sc i} emission feature and this fact
indicates that the latter formed in the circumstellar medium. We
also point out the complex emission-and-absorption profile of
H$\alpha$ (see Fig.\,\ref{H+Na}). Note that none of
our spectra shows the narrow emission lines with low-excitation
metal lines  identified by Kameswara~Rao et al. [\cite{Rao}].

\subsection{Identification and measurement of the parameters of
          spectral features}

According to our estimates, which are close to those of
[\cite{Menz}] and [\cite{Rao}], QY\,Sge has a spectrum of a late F-type
supergiant with more than half of its lines belonging to iron. This fact
allowed us to use the spectrum of the $\alpha$\,Per (Sp=$F5\,Ib$)
supergiant taken with the same instruments as the spectrum of QY\,Sge for
identifying the lines and selecting the least blended among them and to
adopt the laboratory wavelengths necessary for the determination of radial
velocities mostly from the solar-spectrum tables [\cite{Pierce}].

Table~\ref{lines} lists the identifications of the
main lines that we used to study the radial-velocity pattern. The
same table gives for all our  spectra the central residual
intensities ``r'' of the selected absorptions and the radial
velocities inferred from the lower parts of their profiles.
Figure~\,\ref{vel}  gives examples of
dependencies  $V_r(r)$ between these two parameters. The scatter
of data points on these relations allows one to judge about the
real errors of the inferred $V_r$, which include not only
measurement errors, but also uncertainties of the laboratory
wavelengths, effects of unaccounted blending and local
deformations of the dispersion curves employed. Variations of the
scatter from one spectrum to another are due not only to
different exposures, but also to the variations of the profile
shapes with time. Figure~\,\ref{vel} compares the
$V_r(r)$ dependencies obtained for the same spectral range 
from the spectra taken on August 16 , 2003  (when the most
symmetric and narrow metal absorptions were observed) and August
28, 2004. The mean errors of the determination of  $V_r$
estimated from these data based on a single line differ and are
equal to  0.6 and 2.0\,km/s, respectively. In view of the errors
of the determination of  $V_r$, we rounded the corresponding
values in Tables\,\ref{RV} and \ref{lines} to 1\,km/s.

Derivation of the mean velocities for individual dates requires
the allowance for differential line shifts, which, according to
our data, are real and variable: they are barely visible in the
spectrum taken on August 16, 2003, but quite conspicuous in our
other spectra. By their absolute magnitude (about 10\,km/s) the
maximum mutual shifts are close to those pointed out by Kameswara~Rao et al.
[\cite{Rao}], but their behavior differs: instead of
the  sharp rebound of $V_r$ as inferred from resonance lines and
low-excitation absorptions, we observe monotonic increase of
velocity with increasing line intensity (the lower  $V_r(r)$
curve in Fig.\,\ref{vel}). However, such dependences
can also be found by analyzing the data given by Kameswara~Rao et al.
[\cite{Rao}]. Two circumstances corroborate this: (1)
the spectral resolution of [\cite{Rao}] is at least as
good as ours and their signal-to-noise ratio (judging from
Figs.\,1, and 2 adopted from [\cite{Rao}]) is even
higher than our S/N, whereas the errors of the
mean $V_r$ values reported by the above authors are close to our
errors for a single line; (2) absorptions with low potentials of
excitation of the lower level are among the strongest ones.

The slope of the  $V_r(r)$ relation can be affected by the line
asymmetry, which is immediately apparent in the profiles of
Fe{\sc i} lines shown in Fig.\,\ref{profiles}. In the
spectra taken on August 28, 2004 the absorption cores are
redshifted relative to the wings, and are more conspicuous
compared to the profiles of deeper lines. As a result, the
velocity measured at half the central depth (the mean velocity
for the line as a whole) increases with depth to a lesser extent
than the velocity measured from the bottommost part of the
profile (the crosses and squares in Fig.\,\ref{vel},
respectively).

The  two rows of Table\,\ref{RV} give the  $r$
and $V_r$ values (corresponding to the lower parts of the
profiles of photospheric absorptions) for the right-hand ends of
the  $V_r(r)$ curves represented by the lines of the 15th Fe{\sc
i} multiplet. Below, rows 4 and 5 of the same table give the same
quantities for the weakest lines (the left-hand ends of the
$V_r(r)$ curves). The table confirms the conclusion about the
radial-velocity variability of QY\,Sge made by Kameswara~Rao et al.
[\cite{Rao}]. Unfortunately, the above authors did not
analyze  differential line shifts in the spectra of QY\,Sge and
give only the mean velocities  (from $-23$ to $-9$\,km/s)
averaged for each season, which, however, agree well with our
velocities. To illustrate the variable asymmetry of the Fe{\sc
i}\,(15) line, we give in row 3 of Table\,\ref{RV}
the shifts of the data points on the blue and red slopes of the
line profiles relative to the line cores, measured at half of their depth.
 To reduce noise, we
averaged the profiles of several absorptions of similar intensity
and show the result in Fig.\,\ref{profiles}. As is
evident from the figure, the profiles obtained in 2003 and 2004
are not just asymmetric, but appreciably deformed. On both the
red slope of the former profile and the blue slope of the latter
profile averaging confidently reveals depressions similar to
those observed in the spectra of SB2 binaries. However, these
results do not confirm the hypothesis about the binary nature of
the star: it assumes that both the primary and secondary
components of the profile should shift in opposite directions.
However, the  2003 profile is redshifted entirely relative to the
2004 profile. A comparison of the spectra taken in  2002 and 2003
also shows that variations of microturbulent velocity and/or
radial gradient of velocity in the star's atmosphere can be
neither the only nor even the main cause of the variations of
absorption profiles. The increase of the line depth in the first
spectrum is not accompanied by their broadening and
intensification. This is especially apparent in absorptions of
moderate and low intensity ($r \ge 0.7$). Both the specific shape
of the profiles and the nature of its variations  can be
explained by the scattering of the star's radiation by the dust
envelope with inhomogeneous density distribution. Scattering on
dust grains is known to cause the broadening and asymmetry of
photospheric absorptions and their shift toward longer
wavelengths [\cite{VanBlerkom1978}]. Such a redshift
has been known since long, but it could be confidently detected
only after the measurement of the systemic velocities of cool
supergiants from the hydroxyl lines in their spectra (see, e.g.,
[\cite{Wallerstein1977}]). The blue POSS-2 plate image
(http://archive.stsci.edu/dss) of the area surrounding  QY\,Sge
shows a scattering dust nebular (with a size of up to
$30^{''}$), which suggests the pre\-sence of an expanding dust
shell near the star.

\subsection{Peculiar profile of the  Na{\sc i} D lines}\label{Na-profile}

The emission lines (or components) of the sodium doublet can be seen in
the spectra of a number of well-studied post-AGB stars. For example, the
spectrum of the high-latitude supergiant 89\,Her contains a weak emission
feature, which is blended with what appears to be an interstellar
absorption [\cite{Waters}]. A complex emission-and-absorption profile of
Na{\sc i} D lines can also be seen in the very high-resolution ($\approx
10^5$) spectrum of HR\,4049 taken by Bakker et al. [\cite{bakker}] with
the CES spectrograph attached to CAT telescope. Bakker et al.
[\cite{bakker}] suggested that the variability of part of the Na{\sc i}
components is due to variable emission.

We use Na{\sc i} D lines to analyze the structure of the
circumstellar matter in the QY\,Sge system. The local corrector
of the star's position can be tuned so as to make the portions of the
image located at different distances from the center of the
star's image to be projected onto different sections of the image slicer
(each with a size of 0.6$\times$1.8~arcsec). In the case of about 1~arcsec
seeing and image-corrector errors no greater than 0.1~arcsec it is safe to
say that the fractions the flux coming from the star and its neighborhood
differ for different sections. Fig.\,\ref{Na_space} shows the profiles of
Na{\sc i} D lines for three angular distances from the star's center. We
point out three important features of this figure. First, the relative
intensities of narrow emission features remain unchanged after the
normalization of each of the spectra to its own continuum level. The most
simple explanation of this phenomenon consists in the assumption that the
size of the formation region of the narrow lines of the sodium doublet is
smaller than the angular resolution of the telescope. Second, the
intensity ratio of the lines of the doublet is close to unity, whereas in
the absence of self absorption the upper levels are populated
proportionally to their statistical weights and the ratio in question must
be close to two. Equal intensities can be explained either by the high
concentration of sodium atoms when resonance scattering on lines levels
out the intensity ratio of the emission components (in this case the
number of atoms along the line of sight must be no less than
$10^{10}$\,cm$^{-2}$), or by the mechanism of resonance fluorescence,
where the ratio of emission intensities is determined by the ratio of the
illuminating fluxes (in the cores of the lines of the photospheric
spectrum). Fluorescence can be observed only in the case where the
illuminating source (the star) cannot be directly observed and its
radiation does not exceed the weak fluorescence. Third, it is evi\-dent
from the wide (high-velocity) components of the lines of the doublet that
the contribution of blueshifted emission increases and that of redshifted
emission decreases with increasing distance. The enhancement of the
blueshifted emission as we move from the star along the nebula indicates
that we observe the ever increasing relative contribution from the regions
of the circumstellar shell that move toward us, whereas the contribution
of receding regions remains virtually unchanged. This effect can be
explained assuming that the regions where the line-of-sight velocity of
gas is directed ``toward'' and ``away from'' observer are spatially
separated (within a 1.8~arcsec square). If we further assume that
high-velocity motions are axisymmetric (as Kameswara~Rao et al. [\cite{Rao}]
adopt in their model), the data shown in Fig.\,\ref{Na_space} can be used
to estimate the ratio of the opening angle ($\alpha$) of the cones in
which the high-velocity motions are contained and the tilt angle ($\beta$)
of the cone axis to the sky plane: $\beta + \alpha/2 = 90^{o}$. And,
finally, Fig.\,\ref{Na_space} leads us to conclude that profiles of such a
shape can be observed only in the cases where the radiation of the central
star is strongly absorbed, i.e., if the star is partially or totally
hidden from direct observation. For example, R\,CrB-type stars (UW\,Cen
[\cite{Rao2004}],  S\,Aps [\cite{Goswami1997}] and R\,CrB itself
[\cite{Rao2006}]), and FG\,Sge [\cite{Kipper1996}] exhibit simi\-lar profiles of the
resonance sodium doublet during periods of light decline with the
only exception that their high-velocity components are much wider
and overlap, and narrow emissions are also rather wide (about
0.5\,\AA{}).

Let us consider the temporal variations of Na{\sc i} D lines.
Table~\,\ref{RV} lists the parameters of the
emission-and-absorption profile of the resonance Na{\sc i} D2
line in the spectrum of QY\,Sge at all our observing times and
Fig.\,\ref{NaD} shows  its shape and temporal
variations. The $r$ values in Table\,\ref{RV}
correspond to the peaks of the narrow emission feature and to the
highest of the two wide emission humps, and to the core of the
deepest (redshifted) absorption component, whereas the $V_r$
quantities correspond to the upper part of the emission peak and
lower parts of the wide emission and absorption as a whole. The
profile agrees with that described by Kameswara~Rao et al.
[\cite{Rao}] to the last detail. It consists of a wide
emission, which can be seen to extend from $-170$ to $+120$\,km/s
and whose central part is cut by an absorption feature, which, in
turn, is parted in two by a narrow (16\,km/s at $r$\,=\,2.5)
emission peak. The narrow emission and the base of the wide
emission are symmetric. The absorption is asymmetric: it is
abruptly bound on the red side (the entire intensity drop fits
inside the  $0 <V_r<$15\,km/s interval), and the lower part of
the blue wing is also abrupt, but it has a bend at $V_r \approx
-$50\,km/s, the wing becomes flat and extends at least out to
$V_r$ $\approx$ $-$95\,km/s.

We use the $V_r$ values for the absorption as zero points for estimating
the extent of its blue wing. We give these estimates ($\Delta V_r$) in the
same Table\,\ref{RV}. It is evident from Fig.~3 in
the paper by Kameswara~Rao et al. [\cite{Rao}] that the
long-wavelength component of the high-velocity profile becomes stronger
in the spectrum taken in 2000 compared to the spectrum taken in 1999.
Our subsequent observations (Fig.\,\ref{NaD}) confirm this
variability of the spectral profile ignored by the above authors. On the whole,
it is safe to say that we have found conclusive evidence for the temporal variability
of the intensity of the narrow emission and the ratio of the intensities of the
humps of the wide emission, although the positions of the main components of the profile
remain unchanged on the radial-velocity scale.

The fixed position of the  Na{\sc i} emission features implies that they form
in the regions that are external to the photosphere of the supergiant. The main
problem for interpreting the Na{\sc i}\,(1) profile is the narrow emission that
overlaps the absorption core. With this feature ignored, the profile can be
represented as a combination of the P\,Cyg-type wind profile and interstellar
absorption. Photospheric absorption must also be present, however, it cannot
contribute signi\-ficantly: it could only slightly lower the central part of the
wide emission -- by no means to the continuum level -- and, as is evident
from Table\,\ref{RV}, the absorption component of the Na{\sc i}
line reproduces neither the radial velocities of strong Fe{\sc i} photospheric
lines nor their oscillations. As for the P\,Cyg-type profile, here we mean its
type III variant according to Beals [\cite{Beals}]: blueshifted
absorption between two emission humps with the blue hump significantly higher
than the red hump. In our case the latter condition is not satisfied, because
the red component is trimmed and weakened by interstellar absorption.
According to the available data ([\cite{Brand}] et al.), the mean
heliocentric radial velocity of the interstellar medium in the solar
neighborhood at the Galactic latitude of QY\,Sge ($l = 58^{\rm o}$) is equal to
about $-15$\,km/s and lies in the $0 \div 10$\,km/s interval at greater
 distances. The red component of absorptions, which is deeper in
the Na{\sc i}\,(1) profile and dominates in the profile of the K{\sc i}\,(1)
7699\AA{} resonance line,  falls just within this velocity interval.

\subsection{The H$\alpha$ profile}

\begin{figure}[t]
\includegraphics[width=14.0cm]{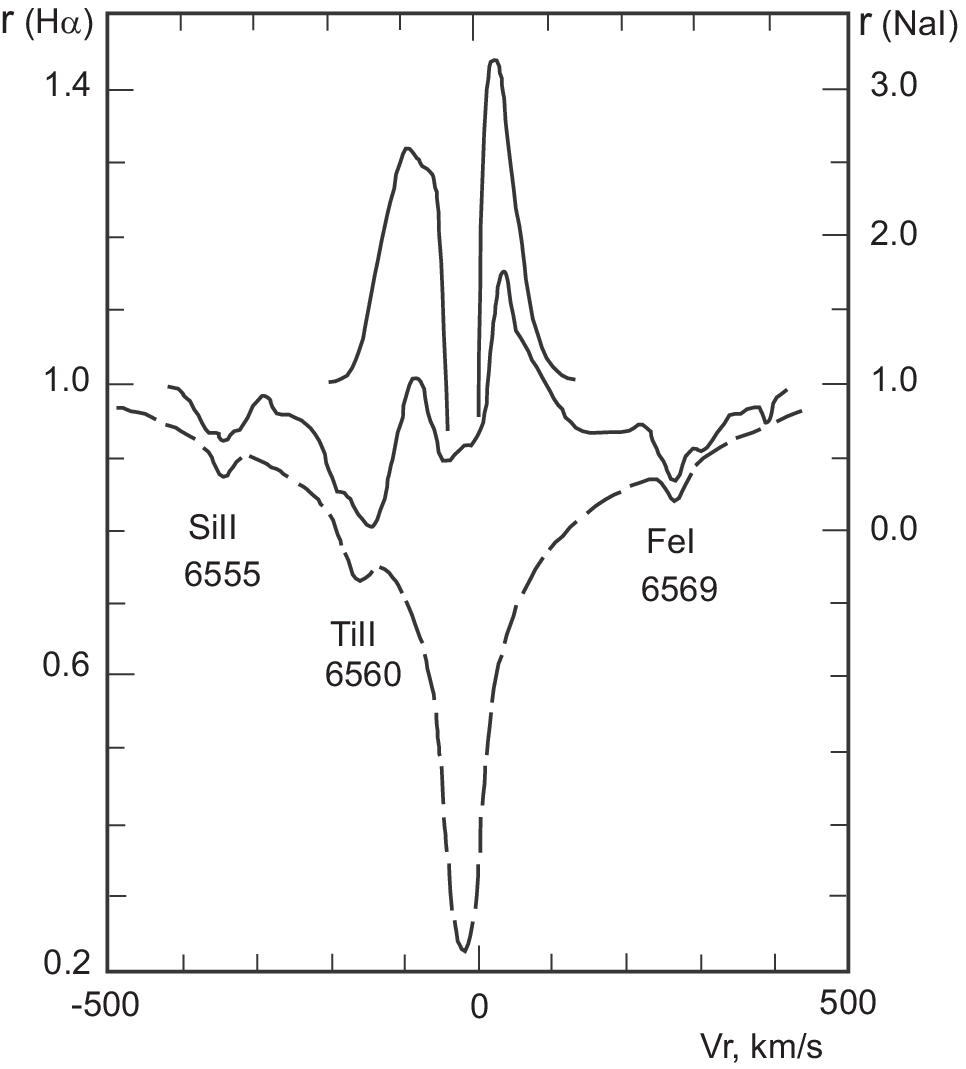}
\caption{From top to bottom: Na{\sc i} 5890\,\AA{} and H$\alpha$ line profiles in the spectrum of
         QY\,Sge taken on August 28, 2004 and the profile of the H$\alpha$
         line in the spectrum of  $\alpha$\,Per. Telluric absorptions and
         the central  Na{\sc i} emission are removed and the profiles are
         smoothed. The $r$ scales for the H$\alpha$ and Na{\sc i} lines
         are shown on the left and right, respectively.}
\label{H+Na}
\end{figure}

\begin{figure}[t]
\includegraphics[width=14cm]{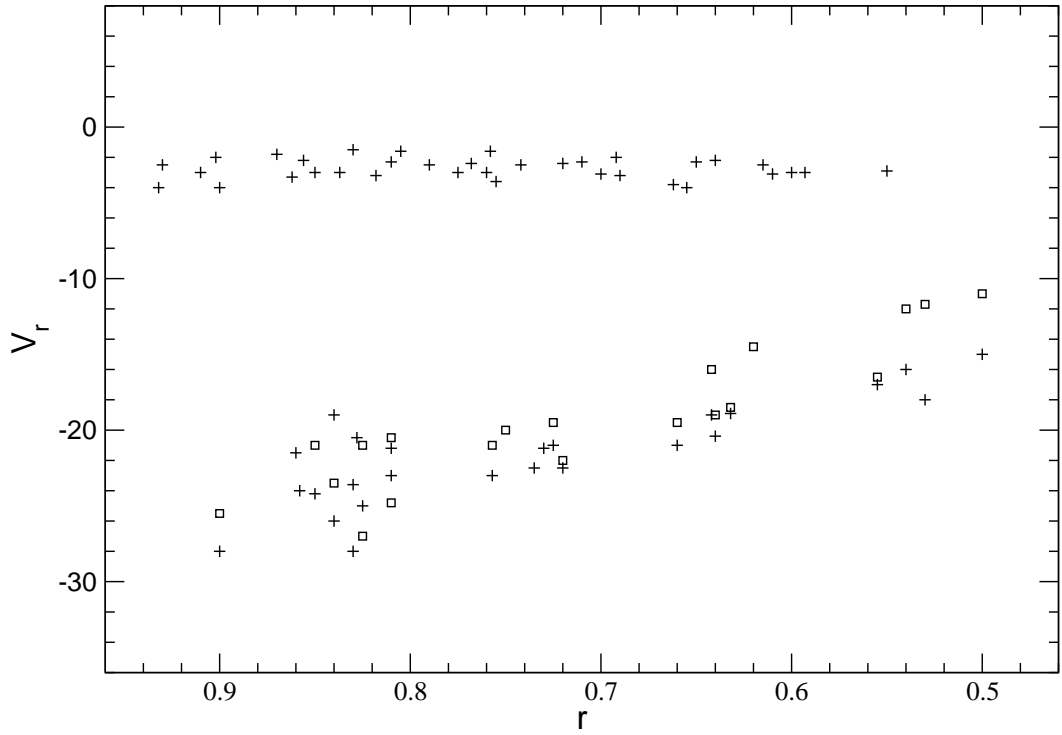}
\caption{Heliocentric radial velocity as a
      function of the central residual line intensity in the spectra
      of  QY\,Sge taken in 2003 (top) and 2004 (bottom, the crosses and
      squares show the $V_r$ values inferred from the entire lines and
      their cores, respectively).}
\label{vel}
\end{figure}

\begin{figure}[h!]
\includegraphics[width=12cm]{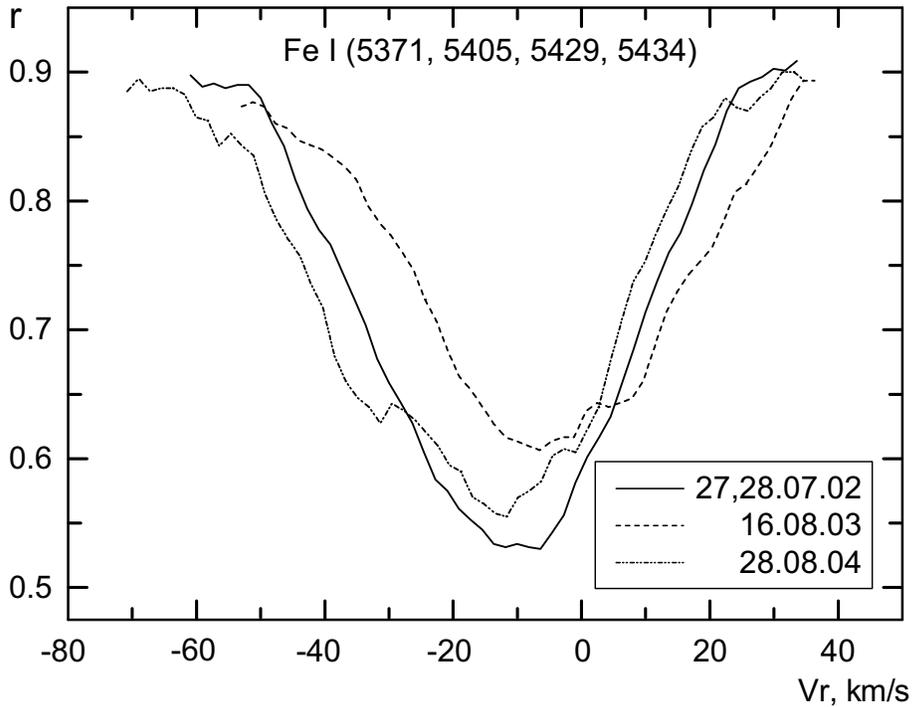}
\caption{Comparison of the averaged profiles of the Fe{\sc i}\,(15)
         group of lines in the spectra taken in  2002, 2003, and 2004.}
\label{profiles}
\end{figure}

\begin{figure}[t]
\includegraphics[width=12cm]{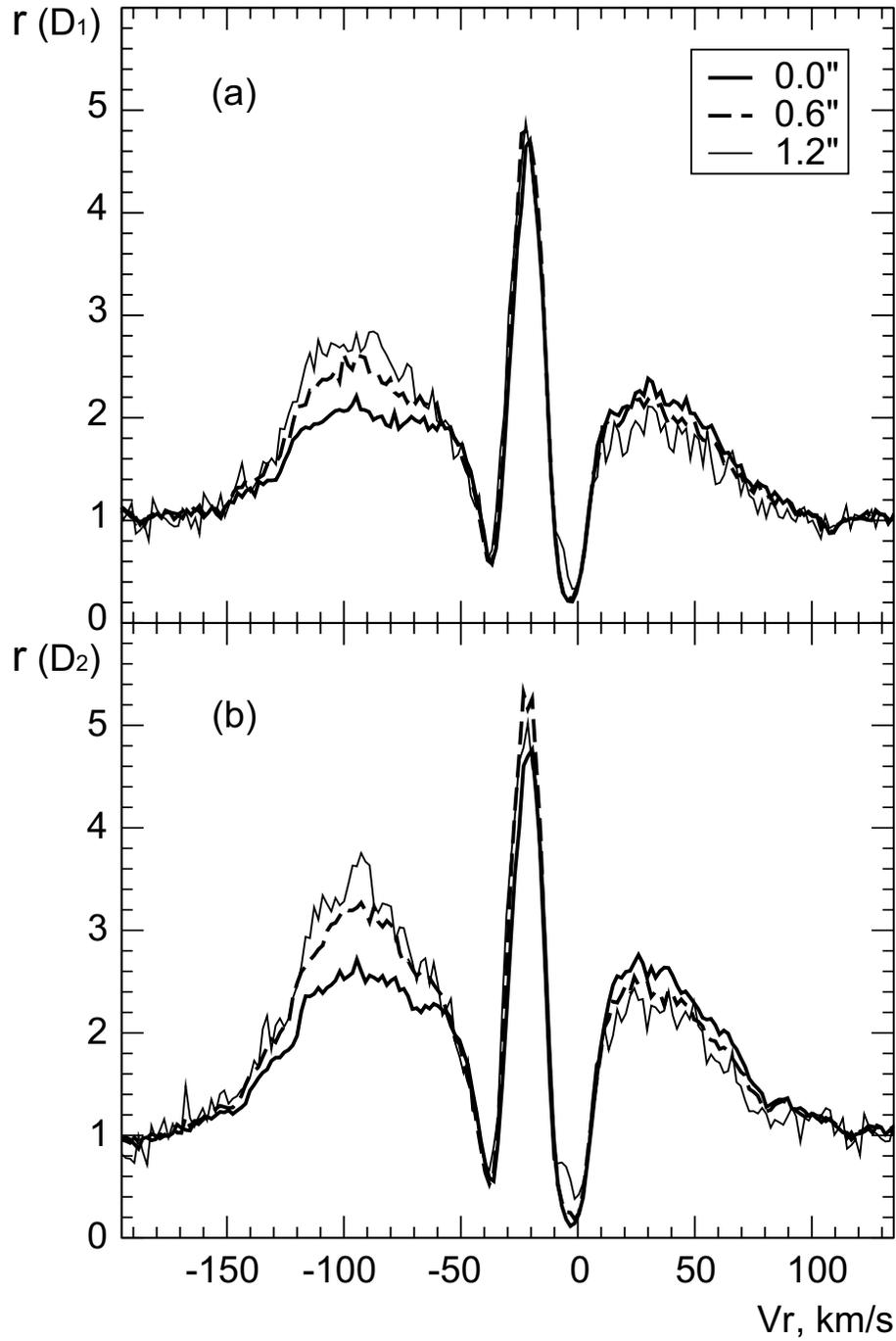}
\caption{Profiles of the Na{\sc i} D1 (top) and D2 (bottom) lines in the
        spectrum of QY\,Sge (observed on August 16, 2003) at different
	 distances from the center of the star's image.}
\label{Na_space}
\end{figure}

\begin{figure}[t]
\includegraphics[width=14cm]{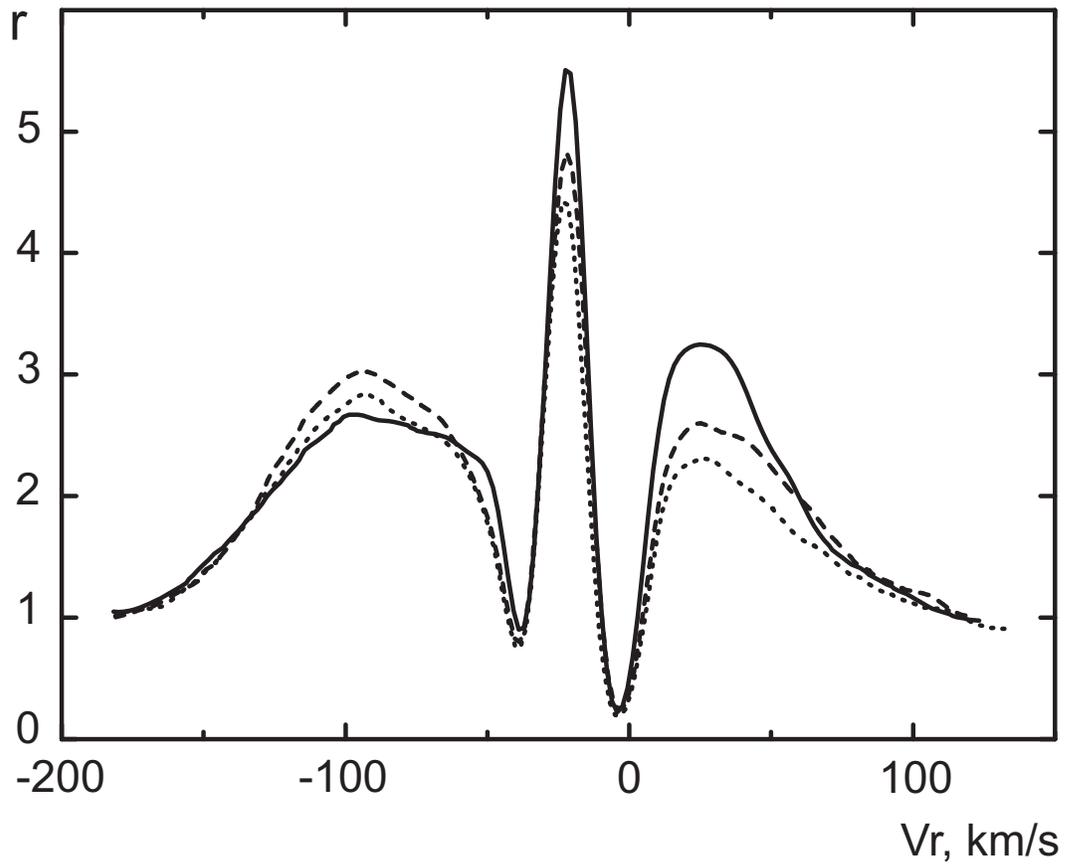}
\caption{Temporal variations of the profile of the Na{\sc i} D2 line
in the spectrum of  QY\,Sge. The dotted, dashed, and solid lines show
the data for 2002, 2003, and 2004, respectively. Telluric absorptions are
removed and the profiles are smoothed.}
\label{NaD}
\end{figure}

Compared to Na{\sc i}\,(1) lines, the H$\alpha$ line in the spectrum of QY\,Sge
is less pronounced, but the resemblance of their profiles is evident.
Unfortunately, Kameswara~Rao et al. [\cite{Rao}] do not report their
H$\alpha$ profile, but only point out that the H$\alpha$ is comparable to the
broad  Na{\sc i}\,(1) emission feature. Our data too corroborate this result,
as is evident from Fig.\,\ref{H+Na}, which shows the profiles of
both lines in the spectrum taken in 2004 along with the H$\alpha$ profile in
the spectrum of the comparison star $\alpha$\,Per $F5 Ib$. H$\alpha$ is weaker
than emission in  Na{\sc i}(1) lines, it almost floods the photospheric
absorption, resulting in a small intensity difference: 0.7$< r < $1.1. This
line is totally indistinguishable in the low-resolution spectra taken by
Menzies and Whitelock [\cite{Menz}]. Photospheric wings, especially
the short-wavelength one, are clearly visible, however, we bear in mind the
problems associated with drawing the continuum of the echelle spectrum and
restrict our analysis to investigating the central, wind-produced, part of the
spectrum. The spectrum of QY\,Sge does not show the narrow emission in
H$\alpha$ (see Fig.\,\ref{H+Na}, it was also removed from the
Na{\sc i} profile), and the positions and relative heights of broad emission
humps repeat themselves in the profiles of the H$\alpha$ and Na{\sc i}
features. The same appears to be true for the positions of the absorptions that
separate the peaks, and small discrepancies between the $V_r$ values listed in
Table\,\ref{RV} may be due to the interstellar component of the
Na{\sc i} feature. It is also possible that the wind-produced absorption in
H$\alpha$ also contributes  (along with the Ti{\sc ii} 6560\,\AA{} line) to the
depression at $V_r \approx -150 \div -200$\,km/s.

\section{Determination of model parameters and chemical abundances}

Absorption lines in the spectrum of  QY\,Sge have a substantial width,
$FWHM\approx$45\,km/s. Such a broadening of the profiles of photospheric absorption
lines is observed in the spectra of a number of post-AGB stars. For example,
the widths of photospheric lines in the spectrum of AFGL\,2688 may be as high
as 40\,km/s  [\cite{Egg1,Egg2}]. The mechanisms of such a
strong line broadening in the spectra of supergiant stars are not yet entirely
understood. Attempts have been made to explain the broadening  in the
spectrum of  AFGL\,2688 by scattering on moving dust [\cite{Egg2}],
when we see only the radiation scattered by dust and the star is hidden by a
dust torus.

Most of the lines in the spectrum of  QY\,Sge are strongly blended because of
their appreciable width, thereby complicating the measurement of their
equivalent widths. We measured most of the equivalent widths $W_{\lambda}$ of
the lines in the  highest signal-to-noise ratio spectrum taken with PFES; in
some cases we measured $W_{\lambda}$ in higher-resolution spectra taken with the
NES spectrograph. The increased spectral resolution gives no advantages for
identifying spectral features, but allows more accurate separation of blends.
In addition, high spectral resolution allows complex profiles with narrow
emissions to be resolved into individual components.

We  apply the method of model atmospheres to numerous iron lines to
determine the parameters of the star's atmosphere. As a first approximation, we
use the atmospheric parameters of QY\,Sge obtained by Kameswara~Rao et
al. [\cite{Rao}]. We determine the effective temperature  $T_{eff}$
from the condition that the iron abundance determined from individual  Fe{\sc i}
lines should be independent of the excitation potential of the corresponding
lines. We fix surface gravity $log\,g$ based on the condition of ionization
balance, i.e., that the iron abundance determined from  Fe{\sc i} lines must be
equal to the iron abundance determined from the lines of  Fe{\sc ii}. We
determine the microturbulence velocity $\xi_t$ from the condition that the iron
abundance must be independent of the equivalent width of the corresponding
Fe{\sc i} lines. In our computations we use Kurucz's [\cite{kurucz}]
grid of models and WIDTH9 program to determine the elemental abundances. We
adopt the excitation potentials and oscillator strengths for all lines, as well
as broadening constants from VALD atomic line database [\cite{vald}].

\begin{figure}[t]
\includegraphics[width=17cm]{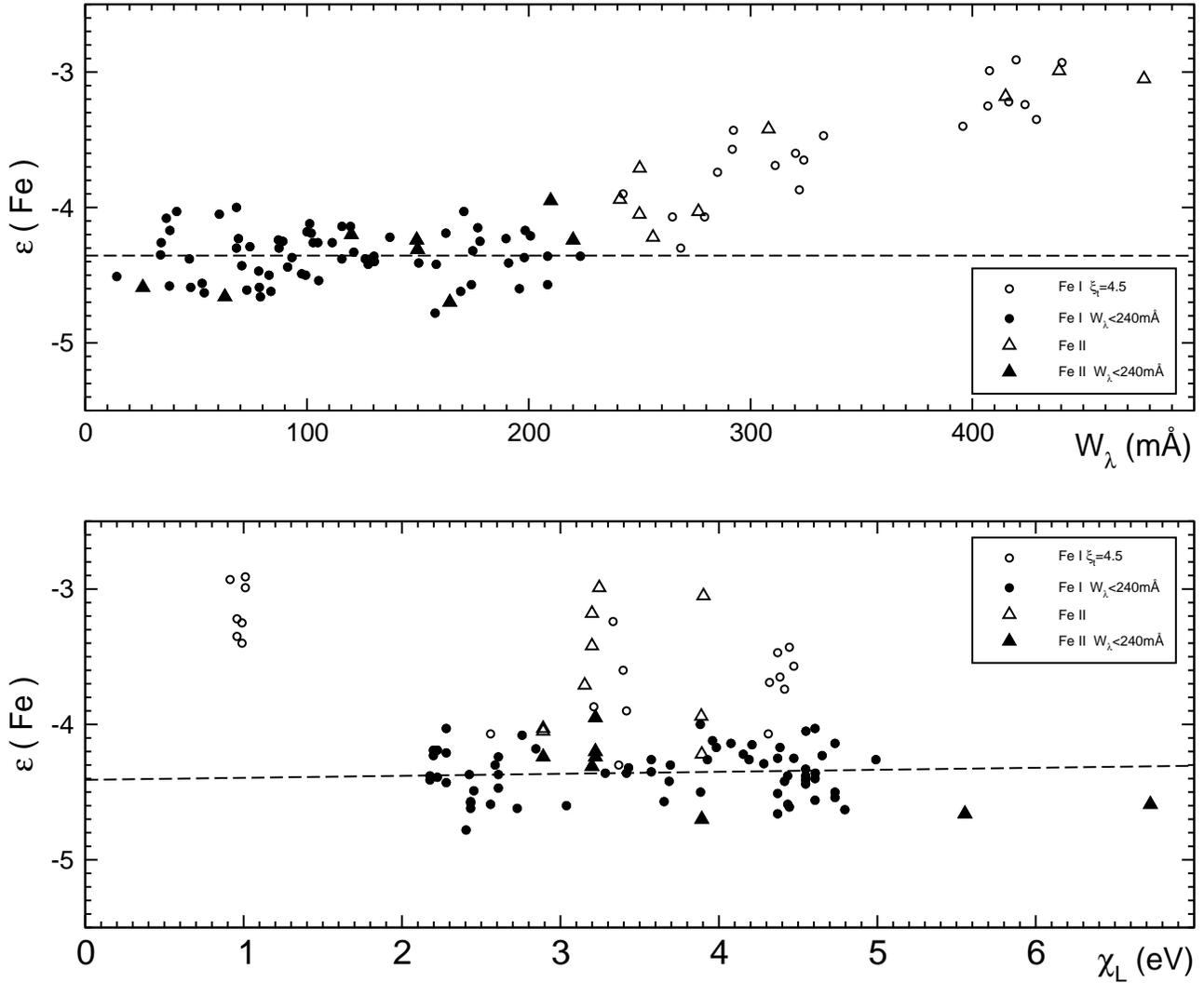}
\caption{Iron abundances, $log\,\epsilon(Fe)$,
determined from  Fe{\sc i} and Fe{\sc ii} lines as a function of: (a)
equivalent line width $W_\lambda$ and (b) excitation potential $\chi_L$ of the
lower level.}
\label{atm_par}
\end{figure}

Like in the case of most of the supergiants, we could not find a
single microturbulence velocity to fit the full set of lines
forming at different depth in the atmosphere of  QY\,Sge.
Figure~\,\ref{atm_par}a shows the iron abundance
$log\,\epsilon$(Fe) as a function of the equivalent width of
individual  Fe{\sc i} lines (circles) at $\xi_t$=4.5\,km/s. As is
evident from the figure, with the adopted parameters of the model
atmosphere the iron abundances determined from weak lines with
equivalent widths $W_\lambda<$240\,m\AA{} (filled circles) are
independent of the line intensity. The dashed line shows the
regression relation based on weak lines. At the same time, strong
lines with $W_\lambda>$240\,m\AA{} (open circles) deviate
significantly toward higher  $log\,\epsilon(Fe)$. The attempts to
determine the microturbulence velocity from strong lines yield
$\xi_t$=15\,km/s. This situation is often explained by the
variation of the microturbulence velocity with depth in the
star's atmosphere. Deviations from local thermodynamic
equilibrium (LTE) in the atmospheres of supergiant stars also
play an important part: strong lines are usually more sensitive to
non-LTE effects. The allowance for deviations from LTE for  Fe{\sc
i} lines makes the variation of microturbulence velocity with
depth somewhat more pronounced [\cite{lubimkov}] and
therefore the increase of  $\xi_t$ for stronger lines cannot be
explained by deviations from LTE in the atmosphere of  QY\,Sge.
This conclusion is indirectly corroborated by a similar result
obtained from  Fe{\sc ii} lines, although the latter are less
sensitive to non-LTE effects. The triangles in
Fig.\,\ref{atm_par} show the results obtained for
Fe{\sc ii} lines: the filled and open triangles correspond to the
lines with $W_\lambda<$\,240\,m\AA{} and $W_\lambda>$240\,m\AA{},
respectively. The microturbulence velocity
$\xi_t$=4.5$\pm$0.5\,km/s that we determined from weak lines
agrees with the results obtained by Kameswara~Rao et al.
[\cite{Rao}].

Figure\,\ref{atm_par}b shows the iron abundance,
$log\,\epsilon$(Fe), as a function of the excitation potential
$\chi_L$ of the lower level of the corresponding transition. The
dashed line shows the result of linear regression for Fe{\sc i}
lines with $W_\lambda<$240\,m\AA{} (filled circles). At
$T_{eff}=6250\pm$150\,K the iron abundance inferred from weak
Fe{\sc i} lines does not correlate significantly with excitation
potential. Our excitation temperature is somewhat higher than the
temperature $T_{eff}$=5850$\pm$200\,K determined by Kameswara~Rao
et al. [\cite{Rao}], however, the discrepancy is
insignificant given the errors of temperature determination.

The iron abundances determined from  Fe{\sc i} and Fe{\sc ii}
lines agree within the measurement errors at the surface gravity
value of $log\,g=2.0\pm0.2$. For the given parameters of the
atmosphere of  QY\,Sge the group of  66 weak Fe{\sc i} lines
yields an average iron abundance of $log\,\epsilon$(Fe)=7.64,
which is 0.14\,dex higher than the solar value. The group of
eight weak Fe{\sc ii} lines yield the same iron abundance,
$log\,\epsilon$(Fe)=7.64. It is evident from
Fig.\,\ref{atm_par}a that for parameters
$T_{eff}=6250$\,K, $log\,g=2.0$, and $\xi_t=4.5$\,km/s of the
model atmosphere of QY\,Sge the iron abundances determined from
Fe{\sc i} and Fe{\sc ii} lines agree throughout the entire range
of equivalent widths. Our result for the surface gravity,
$log\,g$=2.0, differs significantly from the  $log\,g$=0.7
obtained by Kameswara~Rao et al. [\cite{Rao}], and this
fact appears to indicate that the luminosity of the star is lower
than it has been so far believed.

\begin{figure}[t]
\includegraphics[width=14cm]{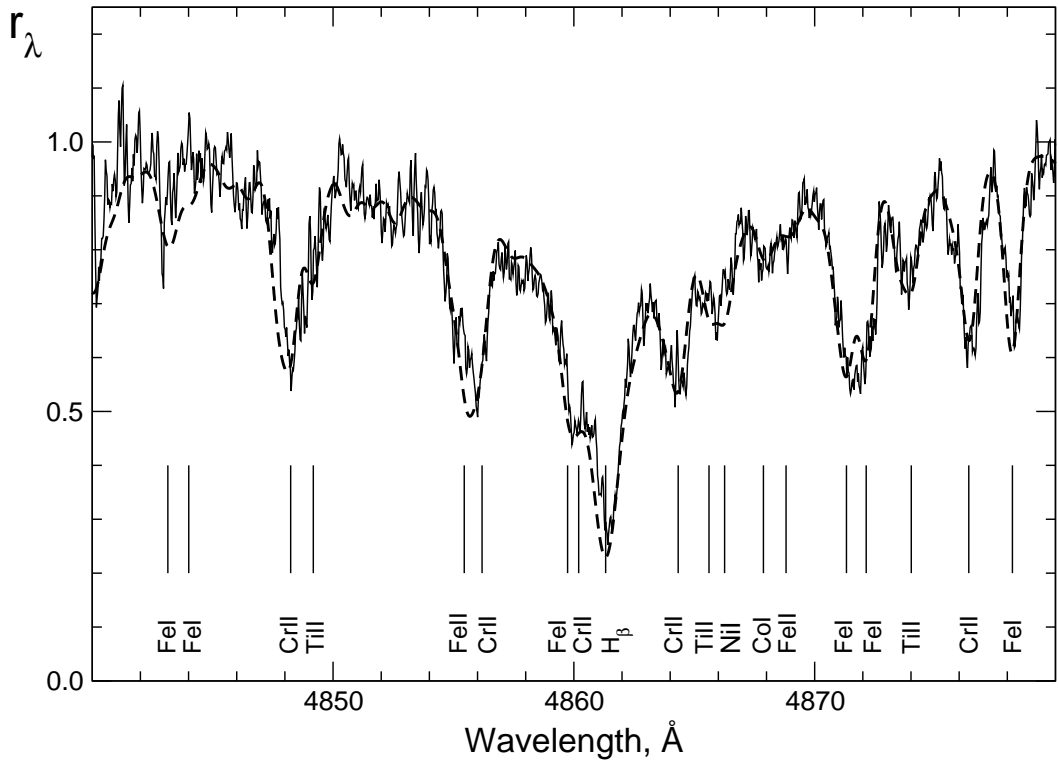}
\caption{Comparison of the observed (the solid line) spectrum of QY\,Sge
        in the region of  H$_\beta$ line and the theoretical spectrum
	computed with parameters $T_{eff}=6250$\,K, $log\,g=2.0$,
	and $\xi_t=4.5$\,km/s and metallicity [Fe/H]=0.}
\label{Hbeta}
\end{figure}

Figure~\,\ref{Hbeta} compares the observed spectrum
of  QY\,Sge in the neighborhood of the  H$_\beta$ line and the
theoretical spectrum computed with our atmospheric parameters
using  {\it SynthV} program [\cite{synth}]. The good
agreement between the two spectra, first, shows that we have
correctly chosen the fundamental parameters of the atmosphere of
QY\,Sge, and, second, disproves the hypothesis that  QY\,Sge
might belong to the class of R\,CrB-type variables characterized
by the underabundance of hydrogen in their atmospheres.

The list of lines with measured equivalent widths and elemental
abundances computed from indivi\-dual lines is available at
http://ales.sao.ru/ftp/pub/ QYSge-lines.html. When computing the
chemical composition of the star we use, like in the case of the
determination of its atmospheric parameters, spectral lines with
equivalent widths $W_\lambda<$240\,m\AA{}.
Table~\,\ref{chem} lists the computed abundances for
various chemical elements in the atmosphere of QY\,Sge averaged
over the entire set of the lines employed. The same table also
gives the standard deviations $\sigma$ of elemental abundances and
the number of lines, ``n'', used in the analysis. The dispersion
serves as a good indicator of the accuracy of observational data
for elements with sufficiently large number of lines. As is
evident from the table, the standard deviation usually does not
exceed $\sigma <$0.27\,dex in the cases where the number of lines
$n>8\div$10. The factors that contribute to the errors of
observational data include, besides the errors of observational
data, the uncertainty of the parameters of the adopted model
atmosphere. Table~\,\ref{error} lists the errors due
to the uncertainty of the main parameters of the star. As is
evident from the table, the large abundance errors arise from the
uncertainty of the inferred effective temperature, especially for
the lines of neutral atoms with low excitation potentials. In the
case of ion lines the uncertainty of surface gravity,  $\log g$,
is the dominating source of error. The uncertainty of
microturbulence velocity contributes only slightly to the
abundance error, because we use sufficiently weak lines with
$W_{\lambda}<$240\,m\AA{}. One must bear in mind the following
two circumstances: first, parameters of the model atmosphere are
not mutually independent  (e.g., a change in temperature entails
a change in surface gravity), and, second, the determination of
relative abundances (with respect to iron) reduces the error due
to the uncertainty of the parameters of the model atmosphere.

\section{Discussion of results}

\subsection{Elemental abundances}

As we pointed out above, the iron abundance in the atmosphere of
QY\,Sge, $[Fe/H]_{\odot}$=+0.14, is somewhat higher than the solar
iron abundance. The overabundance of iron-peak elements is even
higher and reaches its maximum for manganese,
$[Mn/H]_{\odot}$=+0.35, whereas the zinc abundance is almost at
the solar level, $[Zn/H]_{\odot}$=+0.04. The average
overabundance of the iron-peak elements V, Cr, Mn, Fe, Co, Ni,
Cu, and Zn is equal to $[Met/H]_{\odot}$=+0.20. The good
agreement between the V and Cr abundances determined from the
lines of neutral atoms and ions confirms the correctness of our
determination of surface gravity.

We calculate the abundances of two elements -- carbon and
nitrogen -- of the  CNO group. We determine the nitrogen abundance
from the easily identifiable  $\lambda$\,7468\,\AA{} line,
whereas the other neutral nitrogen lines available in the recorded
wavelength region are blended. We find carbon and nitrogen to
be overabundant and the  C/N ratio  in the atmosphere of QY\,Sge
to be close to its solar value. Such a  C/N ratio is indicative
of dredge-up. The carbon abundance decreases as a result of the
dredge-up of the matter processed in the CNO cycle as convection
develops after the star leaves the main sequence. At the same
time, nitrogen abundance increases and therefore the  $C/N$
decreases. During subsequent evolution of the star the carbon
abundance increases and nitrogen is processed into heavier
elements via nuclear synthesis reactions involving $\alpha$
particles. The C/N ratio again increases during the dredge-up of
the matter that participated in helium burning reactions. It is
possible that in the case of  QY\,Sge we observe in the star's
atmosphere the matter processed both during hydrogen burning in
CNO and NeNa cycles and during the $\alpha$-process.

The abundances of elements of the group of light metals are also,
on the average, higher than the corresponding solar abundances.
The $\alpha$-process elements Mg, Si, and Ca are only slightly
overabundant: on the average, $[\alpha/H]_{\odot}$=+0.12, whereas
sulfur is somewhat more overabundant, $[S/\alpha]$=+0.29. The
scandium abundance is equal to its solar value within the
measurement errors. Titanium, on the other hand, is
underabundant, and its abundance determined from  Ti{\sc ii}
lines is much lower than the solar value. Ti{\sc i} lines in the
spectrum of  QY\,Sge are very weak and measurement errors are
comparable to the equivalent widths. We hence consider the result
obtained from  Ti{\sc ii}, namely $[Ti/H]_{\odot}$=$-$0.42, to be
more reliable.

Among the light metals sodium is most overabundant:
$[Na/H]_{\odot}$=+0.89. The sodium overabundance is often explained
by the high sensitivity of the lines of this element to non-LTE
effects. In the spectra of supergiants  Na{\sc i} lines are
usually strongly enhanced compared to the LTE case and therefore
non-LTE corrections to the sodium abundance determined from
individual lines may reach $-0.75$ for supergiant stars
[\cite{Mash}]. We compute the Na abundance in the
atmosphere of QY\,Sge using the weak subordinate lines 5682,
5688, and 6160\,\AA{}, which are less sensitive to non-LTE effects.
If computed with the non-LTE corrections adopted from
[\cite{Mash}] for the parameters of the atmosphere of
QY\,Sge, the sodium abundance is equal to $[Na/H]_{\odot}$=+0.83,
which agrees with the abundance determined in terms of LTE
approximation within the quoted errors. Hence the sodium
overabundance of $[Na/Fe]$=+0.75 observed in the atmosphere of
QY\,Sge is mostly due to the dredge-up of the matter processed
during the NeNa cycle of hydrogen burning.

The abundances of $s$- and $r$-process heavy ele\-ments do not
allow unambiguous interpretation. The $s$-process element Ba is
overabundant relative to iron, $[Ba/Fe]$=+0.33, whereas another
$s$-process element, Nd, is slightly underabundant,
$[Nd/Fe]=-0.18$. One must bear in mind that  Ba{\sc ii} lines are
very strong in the spectrum of QY\,Sge and even the weakest of
these lines located at 5853\,\AA{} -- we use its intensity to
determine the barium abundance -- has the equivalent width of
$W_{5853}$=260\,m\AA{}. Other, even stronger, lines yield even
higher barium overabundance. The $r$-process element Eu is also
slightly overabundant, $[Eu/Fe]$=+0.26. Yttrium forms in the course
of  both $s$- and $r$-processes, however, this element is
substantially underabundant in the atmosphere of  QY\,Sge,
$[Y/Fe]$=$-$0.82.

It is interesting that the atmosphere of  QY\,Sge exhibits strong
underabundance of two elements -- Y and Ti -- characterized by
the highest condensation temperature, $T_{cond}\approx1600$\,K.
We could follow Kameswara~Rao et al. [\cite{Rao}] and
assume that in the case of the star considered we observe only
the effects of condensation of refractory elements. Such effects
are often observed in RV\,Tau-type stars. However, this
hypothesis is inconsistent, first, with the close to solar
zinc-to-scandium abundance ratio, $[Zn/Sc]$=+0.05, whereas the
$T_{cond}$ values for these two elements differ by almost one
thousand degrees. The same is also true for the zinc--calcium
pair, $[Zn/Ca]$=$-0.03$, with equally large difference of
$T_{cond}$. Second, the high iron abundance is also indicative of
the low efficiency of condensation process in QY\,Sge.

\subsection{The radial velocity pattern}

The pattern of radial velocities $V_r$ in the spectra of
RV\,Tau-type stars is rather complex: it includes orbital motion
with periods ranging from several months to several years along
with other types of motions. For example, the orbital period of
AC\,Her -- a typical and well-studied  RV\,Tau-type star --
$P_{orb} \approx 1200^{\rm d}$ [\cite{HVW}] is
complemented by the pulsation component with a characteristic
period of  $P_{puls} \ge 20^{\rm d}$, and also by the likely
precession of the circumstellar disk (torus). It is evident that
long-term homogeneous spectroscopic monitoring needs to be
performed on the time scales of all the instability types
mentioned above.

Our radial-velocity measurements based on the absorption spectrum
of QY\,Sge confirm and extend the conclusion of Kameswara~Rao et
al. [\cite{Rao}] about the variations of $V_r$ both
with time and from one line to another. The available
observational data are still insufficient for directly testing
the hypothesis that QY\,Sge is an RV\,Tau-type object and/or a
binary. At the same time, we gathered new evidence in favor of
the model suggested by Menzies and Whitelock [\cite{Menz}]
 and elaborated by Kameswara~Rao et al. [\cite{Rao}].
 According to this model, a substantial
part of the star's radiation accessible for us (if not the entire
radiation) originates from the inner parts of a toroidal dust
shell. The kinematic situation in the visible part of the torus
determines the width and shift of the line as a whole, whereas
the hypothesis of inhomogeneous dust density distribution is to
be invoked to explain the discrete components of the ``ragged''
profile of the line (see Fig.\,\ref{profiles}).

Although the IR colors of the object are similar to those
observed in OH/IR stars, the radio spectrum of QY\,Sge lacks the
spectral features inherent to these objects
[\cite{Lewis,Squeren}], which would allow
us to determine the systemic velocity. To a first approximation,
the systemic velocity can be determined from the available data
by the position of the narrow emission component of  Na{\sc i}.
We can assume that it forms in the peripheric regions of the
envelope with low velocity gradients and is therefore symmetric,
free of wind-induced deformations, and stationary. However, in
this case two effects are difficult to explain: the variation of
the intensity of the narrow emission component over a one-year
time scale (Fig.\,\ref{NaD}) and the decrease of this
intensity with the distance from the center of the star's image
proportionally to the decrease of continuum flux
(Fig.\,\ref{Na_space}). Within the accuracy of
measurements the velocity inferred from this narrow emission is
equal to $V_r$=$-21.1$\,km/s. We adopt this value as our first
approximation to systemic velocity $V_{sys}$. The conclusion that
the wide  Na{\sc i} emission is radiated by an extended envelope
that spans far beyond the torus follows both from the lower
polarization in this emission compared to the neighboring
continuum [\cite{Trammel}] and the constancy of the
radial velocity determined from the base of the profile. The mean
velocity $V_r$=$-32.4$\,km/s of the broad emission is also stable
within the measurement errors. The assumption that both  Na\,D
emission features form in the circumstellar envelope leads us to
conclude that the envelope is nonuniform.

The model of Kameswara~Rao et al. [\cite{Rao}] leaves a
number of questions unanswered that concern, first, the existence
of emission features whose widths differ by one order of
magnitude and, second, the mutual shift of the broad and narrow
emission components of Na\,D.

Note that the model of the QY\,Sge proposed by Kameswara~Rao et
al. [\cite{Rao}] is not the only one possible. The
peculiarities of the observed optical spectrum mentioned above
can also be explained in terms of a model of a binary system
without a torus. A similar set of spectral features may form in
the extended and inhomogeneous atmosphere of a supergiant star
outflowing via stellar wind. Adopting the model of an optically
thick torus makes it impossible to estimate the distance to
QY\,Sge via standard spectral-and-photometric method, however, it
allows us to avoid the difficulties associated with this method.
We may limit the heliocentric distance to only several hundred
parsecs  [\cite{Rao}] instead of the 9--36\,kpc,
inferred using the method just described by Menzies and
Whitelock [\cite{Menz}].

\subsection{Whether QY\,Sge is an RV\,Tau-type star}

Kameswara~Rao et al. [\cite{Rao}] are inclined to
believe, based on the inferred parameters  $T_{eff}$ and $\log g$
and the detailed analysis of its chemical composition, that
QY\,Sge is an  RV\,Tau-type star. Maas et
al. [\cite{Maas}], who analyzed the chemical composition
of a sample of pulsating RV\,Tau-type stars, later used the
results of Kameswara~Rao et al. [\cite{Rao}] within the
framework of the same criterion. The chemical composition of
QY\,Sge determined from the data obtained by Kameswara~Rao et
al. [\cite{Rao}] indeed resembles the elemental
abundances in the atmospheres of pulsating long-period variables.
Such objects are usually only slightly metal underabundant. Given
that the underabundance of metals is partially due to selective
condensation of metal nuclei onto dust grains, the metallicity
of  RV\,Tau-type stars can be considered to be close to the
metallicity of stars of the Galactic disk. It follows from the
compilation in Table~\,9 from [\cite{Maas}], that for
classic  RV\,Tau-type stars the initial  (i.e., corrected for the
effect of condensation) metallicity is equal, on the average, to
$[Fe/H]_{\odot} \approx -0.5$. The mean ``corrected'' metallicity,
$[Fe/H]_{\odot} \approx -0.2$, determined by Maas et
al. [\cite{Maas}] for a sample of stars with IR excess
(including  QY\,Sge) which in the diagram of IR colors are
located in the domain of  RV\,Tau-type stars, is even closer to
the solar value.

However, the chemical composition similar to that observed in
RV\,Tau-type stars is not unique. The same pattern of elemental
abundances -- weak metal underabundance, nonstandard abundances
of CNO and $\alpha$-process elements, underabundance (in
extremely rare cases, overabundance) of heavy metals -- can be
found among the post-AGB supergiants studied, which do not belong
to RV\,Tau-type stars. Examples include the post-AGB stars
HD\,161796 and HD\,331319 [\cite{hd331319}] or
HD\,133656 [\cite{Winckel}].  RV\,Tau-type stars
display a wide variety of chemical-composition peculiarities,
because they are determined both by the subtype of the Galactic
population they belong to and by the initial mass and subsequent
evolution of each particular star. Earlier,
Wallerstein [\cite{Wall}] also pointed out that
metallicity (and chemical-composition peculiarities as a whole)
of RV\,Tau-type stars is not the fundamental parameter that would
allow a particular star to be attributed to this type. Hence the
metallicity parameter or chemical composition as a whole do not
allow us to classify  QY\,Sge as an  RV\,Tau-type star.

Shocks are known to propagate in the atmospheres of pulsating
stars and the main indicators of these shocks are the blueshifted
emission in hydrogen lines and asymmetric (or split) metal
absorptions [\cite{Gillet}]. This can be illustrated by
the spectrum of AC\,Her. However, the spectrum of QY\,Sge does
not exhibit such features even if taken with a resolution of
$R=60000$, and this fact too casts doubt on the hypothesis that
QY\,Sge may be an RV\,Tau-type star. The fact that the spectra of
RV\,Tau-type stars lack wide  Na{\sc i} and K{\sc i} D-line
emissions provides yet another piece of evidence that is
inconsistent with QY\,Sge being an RV\,Tau-type star.

\subsection{Whether QY\,Sge is an R\,CrB-type star}

Wide emission components of the lines of the sodium doublet
allows us to consider the possibility of including  QY\,Sge into
the so far small group of stars with such a spectral peculiarity
(R\,CrB-type and selected  post-AGB-type stars). A good example
is provided by UW\,Cen, which is an R\,CrB-type star whose
spectrum at minimum light shows a  Na{\sc i} D line profile with
both a broad and narrow emission components
[\cite{Rao2004}]. The broad  Na{\sc i} emission in the
spectrum of  UW\,Cen [\cite{Rao2004}], like in the
spectrum of  QY\,Sge, is blueshifted relative to the narrow
component. There are other examples as well: S\,Aps
[\cite{Goswami1997}], V854\,Cen [\cite{Rao1993}], and Z\,UMi
[\cite{Goswami1999}]. It is important that both these
components show up only when the star is at mi\-nimum light and are
absent at maximum light. Note that emissions in the spectrum of
QY\,Sge  have been observed at all observational epochs over
six years, but no deep minima have been recorded (the only light
variability known is that with a low amplitude and a
period of about 50 days [\cite{Menz}]). Hence, so far
no evidence is available that would be indicative of  QY\,Sge to
satisfy the main criterion of  R\,CrB-type stars
-- sporadic dust ejections resulting in a substantial decrease of the
star's apparent brightness. The second feature  different
for  QY\,Sge- and   R\,CrB-type stars is the Doppler widths of the
high-velocity emission components of the Na{\sc i} doublet. These
components in the spectra of R\,CrB-type stars are variable and
so broad that they mutually overlap [\cite{Rao2006}],
whereas no such behavior is observed in the spectrum of  QY\,Sge.

\begin{figure}[t]
\includegraphics[width=14cm]{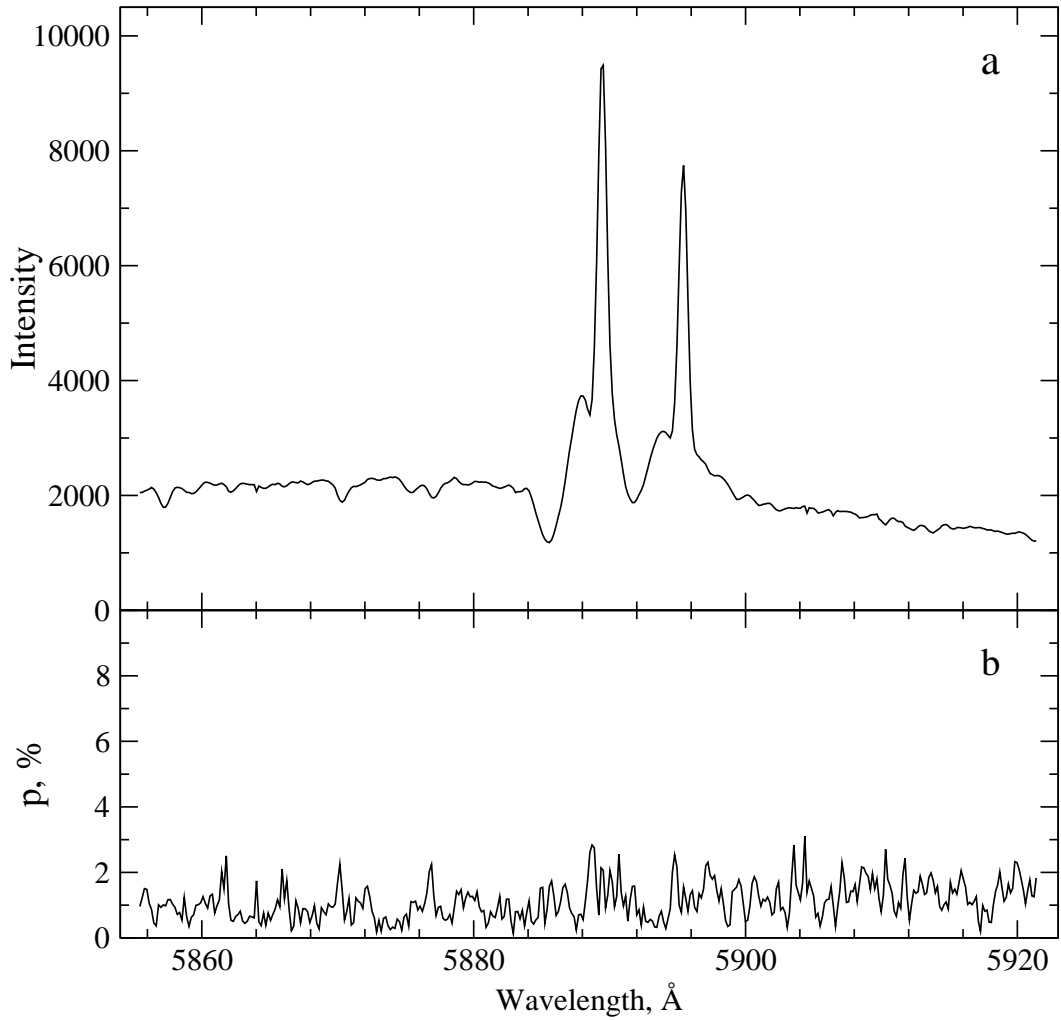}
\caption{Observations of  R\,CrB at minimum light \cite{pfespol}:
         (a) the portion of the spectrum near Na{\sc i}D,
	 (b) the polarization spectrum  P,\,$\%$ in the same wavelength
	 region.}
\label{Polar}
\end{figure}

Third, the pattern of polarization of the emission components
also differs in the two cases. It follows from
spectropolarimetric observations [\cite{Trammel}] made
with a resolution no higher than $R$=1000, the spectrum of
QY\,Sge shows a decrease of the degree of linear polarization
down to the interstellar level (emission becomes absorption as one
passes from the energy distribution to the distribution of
polarization degree). Such a reversal indicates that in the case
of  QY\,Sge the Na{\sc i} emission forms outside the medium that
polarizes the radiation of the photosphere. R\,CrB shows a
totally different pattern. We observed this object with the
spectropolarimeter of the primary focus of the 6-m telescope of
 SAO RAS [\cite{pfespol}] with a resolution of
$R$=15000 when the star was in a state close to minimum light.
Figure~\,\ref{Polar} reproduces the figure from
[\cite{pfespol}], implying that broad and narrow
components of the  Na{\sc i} resonance doublet in the spectrum of
R\,CrB disappeared in the wavelength dependence of the
polarization degree, and polarization degree (1.5--2\%) in the
vicinity of the doublet is the same as in other portions of the
spectrum. Note that before our observations
[\cite{pfespol}] only the polarization characteristics
of broad emissions were known, whereas narrow emissions were not
observed because of the insufficient spectral resolution.
Figure~\,\ref{Polar} leads us to conclude that the
polarizing factor is located between the observer and the
formation regions of the spectra of  E2+BL (classification of the
spectra of stars of R\,CrB type  according to
Clayton [\cite{Clayton1996}]). Hence QY\,Sge differs
from R\,CrB also in spectropolarimetric characteristics of the
sodium resonance doublet. Moreover, as we pointed out above, the
H$\beta$ line in the spectrum of  QY\,Sge corresponds to normal
hydrogen abundance (see Fig.\,\ref{Hbeta}), and this
fact also disproves the  R\,CrB-type classification of QY\,Sge.

\subsection{Comparison with V510\,Pup}

Let us now compare the spectral peculiarities of  QY\,Sge with
the spectrum of the  post-AGB star V510\,Pup (it is the optical
component of the IR source IRAS\,08005-2356), which was also
found to exhibit an emission feature in Na\,D
[\cite{V510Pup}]. Despite this similarity, the spectra
of  QY\,Sge and V510\,Pup differ substantially and their
differences, which are evidently due to the differences between
the structure and geometry of the two systems. As is evident
from Figs.\,\ref{Na_space} and \ref{NaD},
the Na\,D emission in the spectrum of  QY\,Sge is extremely
strong and exceeds the continuum level by a factor of four to
five. The intensity of the emission in the spectrum of V510\,Pup
does not exceed  1.5 continuum intensities. This difference is of
no fundamental importance, because it can be explained by
different degree of shielding of the radiation of the central
star. Unlike  QY\,Sge, spectropolarimetric observations of
V510\,Pup do not show the decrease of polarization in the  Na\,D
lines, at least in low-resolution polarization spectra
[\cite{Trammel}].

Trammel et al. [\cite{Trammel}] conclude that Balmer
emissions in the spectrum of V510\,Pup form in the vicinity of
the star before the scattering on the dust component. The
velocities measured from  Na{\sc i}~D emissions and hydrogen lines
for V510\,Pup agree with each other [\cite{V510Pup}],
allowing us to preliminary conclude that the  Na\,D emission
also forms near the star. The same conclusion can be made for
QY\,Sge: although the absorption profile of  H$\alpha$ is almost
completely flooded by the emission, spectropolarimetry does not
reveal unpolarized H$\alpha$ emission. Note also that the pattern
of the variation of the polarization degree over a wide
wavelength region also differs for these stars. In the spectrum
of  V510\,Pup the polarization degree decreases with increasing
wavelength [\cite{Trammel}], and this fact is
interpreted as the domination of scattered radiation in the blue
part of the spectrum and partial transparency in the red emission
of the disk (torus) that hides the star. In the spectrum of
QY\,Sge the polarization degree increases with wavelength,
suggesting higher scattering multiplicity at blue wavelengths and
disk opaqueness at red wavelengths [\cite{Trammel}]. On
the whole, the conclusions based on the analysis of the
polarization pattern are consistent with the results of the
comparison of the intensities of  Na\,D: in the case of QY\,Sge
we appear to observe the star only indirectly, and the
contribution of high-velocity components of the shell dominates.
The absence of polarization in Na{\sc i} lines in the spectrum
of  V510\,Pup can be interpreted as being due to the low
contribution of emission compared to the radiation of the
photosphere. This contribution is undetectable by low-resolution
spectropolarimetry. Hence the formation regions of the Na\,D
emission may have the same geometry in  QY\,Sge and V510\,Pup,
but the  localization of the Balmer emission near the
V510\,Pup star does not imply the same localization of  the Na\,D
emission.

However, one must admit that the emission feature in  Na\,D
cannot be viewed as indicative of the object belonging to a
certain type of stars or to a fixed stage of evolution. Recall
that a very strong and variable emission component of  Na{\sc i}~D
lines is observed in the spectra of such different stars as the
unique star FG\,Sge [\cite{Kipper}], which has lost
virtually all its atmosphere in the process of evolution, and the
B[e] star CI\,Cam [\cite{Mirosh,Hynes}],
which is the optical component of an X-ray transient. Emission in
resonance line is rather indicative of the gas-and-dust
circumstellar medium with a set of physical parameters over a
wide interval. However,  if combined with spectropolarimetric
observations, the data on the high-velocity motions in the
vicinity of the star as inferred from the  Na{\sc i}~D resonance
doublet allows one to test the hypotheses concerning the geometry
of the system.

\section{Conclusions}

The results of repeated spectroscopic observations (with the resolutions
of $R$=15000 and 60000) of the yellow supergiant QY\,Sge (the IR source
IRAS\,20056+1834) made with the 6-m telescope of SAO RAS lead us to
conclude that the radial velocities measured from lines forming in the
photosphere are variable. We revealed differential line shifts amounting to
10\,km/s.

The complex emission-and-absorption profile of Na{\sc i}\,D lines
invariably contains a very wide emission component (it extends from $-170$
to +120\,km/s). The wide emission is cut at its central part by an
absorption feature, which, in turn, is parted into two by an overlapping
(16\,km/s at $r$\,=\,2.5) emission peak. The positions of Na{\sc i}\,D
emission features remain unchanged and this fact indicates that they form
in the regions that are external to the photosphere of the supergiant. The
$V_r$ dataset allows setting the systemic radial velocity equal to
$V_r$=$-21.1$\,km/s, which corresponds to the position of the narrow
emission component of Na{\sc i}\,D.

Emission in the H$\alpha$ line floods the photospheric absorption
almost completely.

The pattern of the variation of the profiles of emission and
absorption lines and radial velocities measured from different
details of the profiles is consistent with the model of a toroidal
dust envelope that hides the central source and bipolar cones
filled with high-velocity gas. Both in the emission and
absorption features we revealed details indicative of the spatial
and temporal inhomogeneities in the dust and gaseous components
of the object.

Absorption lines in the spectrum of QY\,Sge have a considerable width
$FWHM\approx45$\,km/s, which complicates substantially the analysis of the
chemical composition. We used the method of model atmospheres to determine
the following parameters: effective temperature $T_{eff}=6250\pm150$\,K,
surface gravity $log\,g$=2.0$\pm0.2$, and microturbulence velocity
$\xi_t$=4.5$\pm$0.5\,km/s. The chemical composition of the star's
atmosphere differs insignificantly from the solar composition: we find the
star's metallicity to be somewhat higher than the solar value with the
mean overabundance of iron-peak elements V, Cr, Mn, Fe, Co, Ni, Cu, and Zn
equal to $[Met/H]_{\odot}$=+0.20. We also found carbon and nitrogen to be
slightly overabundant -- $[C/Fe]$=+0.25 and $[N/Fe]$=+0.27 -- with the
C/N ratio close to its solar value. The $\alpha$-process elements Mg, Si,
and Ca are slightly overabundant and their average relative abundance is
equal to $[\alpha/H]_{\odot}$=+0.12, while sulfur is more overabundant,
$[S/\alpha]$=+0.29. We found strong sodium overabundance,
$[Na/H]_{\odot}$=+0.75, which appears to be due to the dredge-up of the
matter processed in the NeNa--cycle. The abundances of heavy $s$-process
elements are lower than the corresponding solar abundances.

On the whole, so far the observed properties of QY\,Sge do not
give grounds to consider it as an  R\,CrB-- or RV\,Tau--type star.
We point out the following especially important aspects besides the
evident need for continuing the spectroscopy of QY\,Sge on a more regular
basis:
\begin{itemize}
\item{} the S/N ratio must further be increased while maintaining
        the already achieved spectral resolution in order to more confidently
        outline the details of complex and variable profiles of absorption lines;
\item{} it is desirable to use high-resolution spectropolarimetry, which
        would, in particular, help to refine the locations of the formation
        regions of individual components of the emission-and-absorption
        Na{\sc i}\,D lines and other features;	
\item{} extremely valuable data on the geometry and kinematics of the
        system can be provided by (high angular resolution) spectra of
        sufficiently bright peripheric regions of the dust envelope, which make it
        possible to ``come close'' to the object and see its central part at
        different angles.
\end{itemize}

\section*{Acknowledgments}

This work was supported by the program ``Observed manifestations of the
evolution of the chemical composition of stars and of the Galaxy'' funded
by the Presidium of the Russian Academy of Sciences, the program
``Extended objects in the Universe'' funded by the Division of Physical
Sciences of the Russian Academy of Sciences, and by the Civil Research and
Development Foundation (CRDF, project RUP1--2687--NA--05). The work is
supported by the Russian Foundation for Basic Research (project
no.~07-02-00247). Our research made use of SIMBAD and VALD databases and
ALADIN interactive sky atlas of the Strassbourgh CDS.

\begin{table}[btp]
\caption{Log of observations of  QY\,Sge, residual intensities
        ``r'' and heliocentric radial velocities for different groups of
         lines. Uncertain values are printed in italics}
\begin{tabular} {l|r|r|r|r|r}
\hline
 Date        &  10.07.98 &   14.07.98 &  27-30.07.02 &  16.08.03 & 28.08.04  \\
Spectrograph &    PFES &  NES      &    NES      &   NES    &   NES     \\
$\Delta\lambda$, nm &400--770&   500--590&480--670&530--660 & 530--660    \\
\hline
 \underline{FeI\,(15)} &            &           &    &   &      \\
 $r$     &    $-$    &    0.52     &   0.53       &    0.61  &    0.56   \\
 $V_r$   &     --9   &   --8     &   --10      &   --3   &    --13  \\
$\Delta V_r$&  $-$    &{\it  --27}, {  \it+29}&--27, +24  &--23, +27& --34, +24 \\
$r_{max}$&    0.90    &  $-$      &  0.86        & 0.93     &  0.86    \\
$V_r$    &   --21   &  $-$      &  --14       &--3      &  --25   \\
\underline{D2, Na{\sc i}\,(1)} &  &              &          &   &       \\
 narrow emission&&&&& \\
  $r$    &      3.2    &     5.7    &    4.9       &   5.0    &    5.9    \\
  $V_r$  &     --22    &   --21    &    --21      &   --21   &  --21    \\
  wide emission&&&&&\\
  $r$    &    2.8 &  2.8 &  2.8    &  3.0 &   3.2    \\
  $V_r$  & --33   &  {\it --31}&  --33  &   --32 &   --33    \\
absorption&             &           &              &          &           \\
  $r$    &      1.6    &      0.4   &      0.1     &    0.1   &     0.1   \\
  $V_r$  &{\it --18} &  {\it --17}&  {--17} &  {--17} &  --18    \\
 $\Delta V_r$&  --  &  {\it --70}&  {\it --80}   &  {\it --75}&  {\it --80}   \\
\underline{H$\alpha$}&             &              &          &          &     \\
  absorption&&&&&\\
 $r$     &     0.73    &    $-$    &    0.75      &    0.79  &   0.89    \\
 $V_r$   &    --23    &    $-$    & {\it --17}  &{\it --13}&  --30   \\
\hline
\end{tabular}
\label{RV}
\end{table}

\clearpage
\begin{center}
\tablecaption{Identification of the spectrum of  QY\,Sge and
         measurements of the central residual intensity  $r$ and
         heliocentric radial velocity ${\rm V_r}$, km/s, based on
         different spectral features.
         See text for explanation. Uncertain values are listed
	 in italics}
\tablehead{\hline  &  &\multicolumn{2}{c|}{1998}&\multicolumn{2}{c|}{2002}&\multicolumn{2}{c|}{2003}&\multicolumn{2}{c}{2004}\\
          Line&$\lambda$&r&$V_r$&r&$V_r$&r&$V_r$&r&$V_r$\\
[2pt] \hline}
\tabletail{\hline \rule{0pt}{0pt} 1& 2 & 3 & 4 &5 &6 &7 &8 &9 & 10 \\ \hline}
\begin{supertabular}{l|c|c|r @{\quad}|c|r @{\quad}|c|r @{\quad}|c|r @{\quad}}
MnI(2)   & 4030.75 &  0.28      &${\it -18}$ &     &           &      &           &     &           \\
MnI(2)   & 4034.48 &  0.43      &${\it -13}$ &     &           &      &           &     &           \\
H$\delta$& 4101.74 &  0.20      &${\it -18}$ &     &           &      &           &     &           \\
TiII(105)& 4163.65 &  0.62      &${\it -12}$ &     &           &      &           &     &           \\
SrII(1)  & 4215.52 &  0.41      &${\it -14}$ &     &           &      &           &     &           \\
FeI(152) & 4222.21 &  0.64      &${\it -10}$ &     &           &      &           &     &           \\
FeI(152) & 4235.93 &  0.39      &  $-11$     &     &           &      &           &     &           \\
CrI(1)   & 4254.33 &  0.59      &${\it -5}$  &     &           &      &           &     &           \\
CrI(1)   & 4274.80 &${\it 0.50}$&${\it -4}$  &     &           &      &           &     &           \\
CaI(5)   & 4318.65 &  0.79      &  $-12$     &     &           &      &           &     &           \\
H$\gamma$& 4340.47 &  0.23      &${\it -12}$ &     &           &      &           &     &           \\
FeI(41)  & 4404.75 &  0.38      &  $-8$      &     &           &      &           &     &           \\
FeI(350) & 4476.04 &  0.74      &  $-13$     &     &           &      &           &     &       \\
FeI(68)  & 4494.57 &  0.58      &  $-13$     &     &           &      &           &     &       \\
FeII(38) & 4508.28 &  0.67      &  $ -9$     &     &           &      &           &     &       \\
FeII(37) & 4520.22 &  0.63      &  $ -9$     &     &           &      &           &     &       \\
BaII(1)  & 4554.03 &  0.55      &${\it -4}$  &     &           &      &           &     &       \\
FeII(38) & 4576.33 &  0.70      &   $-9$     &     &           &      &           &     &       \\
CrII(44) & 4588.20 &  0.71      &${\it -11}$ &     &           &      &           &     &       \\
TiII(50) & 4589.95 &  0.75      &   $-8$     &     &           &      &           &     &       \\
FeI(39)  & 4602.94 &  0.72      &  $-11$     &     &           &      &           &     &       \\
FeII(37) & 4629.33 &  0.70      &  $-10$     &     &           &      &           &     &       \\
FeI(820) & 4643.46 &  0.86      &  $ -9$     &     &           &      &           &     &       \\
MgI(11)  & 4702.99 &  0.70      &${\it -19}$ &     &           &      &           &     &           \\
FeII(43) & 4731.47 &  0.67      &  $-11$     &     &           &      &           &     &       \\
FeI(38)  & 4733.59 &  0.77      &  $ -9$     &     &           &      &           &     &       \\
FeI(554) & 4736.78 &  0.68      &  $-10$     &     &           &      &           &     &       \\
MnI(16)  & 4754.04 &  0.77      &  $ -9$     &     &           &      &           &     &       \\
FeI(821) & 4768.33 &  0.81      &  $-10$     &     &           &      &           &     &       \\
TiII(92) & 4779.98 &  0.79      &  $-15$     &     &           &      &           &     &       \\
MnI(16)  & 4783.42 &  0.78      &${\it -7}$  &     &           &      &           &     &       \\
TiII(92) & 4805.09 &  0.75      &${\it -13}$ &     &           &      &           &     &       \\
H$\beta$ & 4861.33 &  0.35      &  $-9$      & 0.30&  $-10$    &      &           &     &       \\
FeI(318) & 4878.20 &  0.72      &${\it -11}$ & 0.65&  $-14$    &      &           &     &       \\
FeI(318) & 4920.50 &  0.63      &  $-12$     & 0.50&   $-7$    &      &           &     &       \\
FeII(42) & 4923.92 &  0.54      &   $-5$     & 0.39&   $-9$    &      &           &     &           \\
CI(13)   & 4932.05 &  0.84      &  $-17$     &     &           &      &           &     &       \\
BaII(1)  & 4934.08 &  0.55      &${\it -13}$ & 0.43&   $-9$    &      &           &     &       \\
FeI(687) & 4950.11 &  0.83      &  $-12$     &     &           &      &           &     &       \\
FeI(687) & 4966.09 &  0.75      &  $-17$     &     &           &      &           &     &       \\
FeI(984) & 4973.11 &  0.88      &  $-13$     & 0.80&${\it -13}$&      &           &     &       \\
FeI(966) & 4978.60 &  0.83      &  $-13$     &     &           &      &           &     &       \\
NiI(112) & 4980.18 &  0.83      &  $-15$     & 0.74&  $-14$    &      &           &     &       \\
FeI(16)  & 5012.07 &  0.71      &  $-12$     & 0.56&   $-5$    &      &           &     &       \\
FeII(42) & 5018.44 &  0.56      &${\it -21}$ & 0.36&  $-12$    &      &           &     &       \\
ScII(23) & 5031.02 &  0.81      &  $-14$     &     &           &      &           &     &       \\
FeI(114) & 5049.83 &  0.70      &  $-16$     &     &           &      &           &     &       \\
FeI(1)   & 5060.08 &  0.70      &  $-14$     & 0.65&  $-11$    &      &           &     &           \\
FeI(383) & 5068.76 &            &            & 0.70&  $-10$    &      &           &     &       \\
FeI(1094)& 5074.75 &  0.80      &${\it -9}$  &     &           &      &           &     &       \\
FeI(1)   & 5110.41 &  0.50      &${\it -9}$  & 0.42&  $-10$    &      &           &     &       \\
FeI(16)  & 5127.36 &  0.69      &${\it -6}$  & 0.56&   $-5$    &      &           &     &       \\
FeI(1089)& 5162.27 &  0.75      & $-14$      &     &           &      &           &     &       \\
FeI(1)   & 5166.28 &${\it 0.5}$ &  $-9$      & 0.49&   $-9$    &      &           &     &       \\
MgI(2)   & 5167.32 &${\it 0.5}$ &  $-2$      & 0.47&${\it -1}$ &      &           &     &       \\
FeII(42) & 5169.03 &${\it 0.5}$ & $-14$      & 0.35&   $-9$    &      &           &     &       \\
MgI(2)   & 5172.68 &  0.55      & $-10$      & 0.50&   $-6$    &      &           &     &       \\
MgI(2)   & 5183.61 &  0.52      & $-2$       & 0.44&   $-3$    &      &           &     &       \\
FeI(66)  & 5202.33 &  0.78      &${\it -9}$  & 0.69&  $-13$    &      &           &     &       \\
FeI(383) & 5232.94 &            &            & 0.56&  $-10$    &      &           &     &           \\
FeII(49) & 5234.62 &  0.70      & $-13$      &     &           &      &           &     &           \\
FeI(1)   & 5247.06 &  0.61      &${\it -12}$ &     &           &      &           &     &       \\
FeII(49) & 5254.93 &  0.64      & $-10$      & 0.55&  $-10$    &      &           &     &           \\
FeI(383) & 5281.79 &  0.74      & $-12$      & 0.72&  $-12$    & 0.78 &${\it -4}$ &     &           \\
FeI(929) & 5288.53 &  0.93      &${\it -17}$ &     &           &      &       &     &       \\
FeI(553) & 5302.30 &            &            &     &           & 0.76 &  $-3$     &     &       \\
CrII(24) & 5305.86 &  0.84      & $-12$      & 0.78&  $-16$    & 0.83 &${\it -3}$ &     &       \\
CrII(43) & 5313.58 &            &            &     &           & 0.81 &  $-2$     &     &       \\
FeII     & 5316.65 &  0.66      &${\it -6}$  & 0.52&   $-9$    & 0.55 &  $-3$     & 0.56&${\it -17}$\\
FeI(553) & 5324.18 &  0.85      &${\it -10}$ & 0.64&  $-13$    & 0.69 &  $-3$     & 0.55&${\it -20}$\\
FeI(37)  & 5341.03 &            &            & 0.61&  $-12$    & 0.65 &${\it -5}$ &     &           \\
FeI(1062)& 5353.38 &            &            &     &           &      &           & 0.78&  $-19$    \\
FeII(48) & 5362.86 &  0.75      &${\it -16}$ & 0.59&  $-12$    & 0.65 &  $-2$     & 0.58&  $-15$    \\
FeI(1146)& 5367.47 &            &            &     &           & 0.74 &${\it -4}$ &     &           \\
FeI(15)  & 5371.48 &  0.62      & $-12$      & 0.52&   $-8$    & 0.60 &  $-3$     & 0.50&  $-15$    \\
FeI(15)  & 5397.12 &  0.64      &  $-9$      & 0.53&  $-10$    & 0.65 &  $-4$     &     &           \\
FeI(15)  & 5405.77 &  0.64      & $-10$      & 0.54&   $-9$    & 0.61 &  $-3$     & 0.53&  $-12$    \\
CrII(23) & 5420.93 &            &            &     &           &      &           & 0.72&${\it -25}$\\
FeI(15)  & 5429.70 &  0.64      &  $-8$      & 0.52&   $-9$    & 0.60 &  $-3$     & 0.54&  $-14$    \\
FeI(15)  & 5434.53 &  0.68      &  $-7$      &     &           & 0.64 &  $-2$     & 0.56&${\it -17}$\\
FeI(1144)& 5441.34 &            &            &     &           & 0.93 &${\it -3}$ &     &           \\
FeI(1163)& 5445.05 &            &            &     &           &      &           & 0.74& $-21$     \\
FeI(15)  & 5455.61 &            &            & 0.57&${\it -9}$ &      &           &     &           \\
FeI(15)  & 5497.51 &  0.76      & $-12$      & 0.68&  $-10$    & 0.69 &  $-2$     & 0.65& $-19$     \\
FeI(15)  & 5506.78 &            &            &     &           &      &           & 0.67& $-17$     \\
CrII(50) & 5508.62 &            &            &     &           & 0.82 &${\it -3}$ &     &           \\
ScII(31) & 5526.82 &            &            &     &           & 0.74 &  $-2$     &     &           \\
MgI(9)   & 5528.40 &  0.72      & $-12$      & 0.64&${\it -14}$& 0.66 &  $-4$     & 0.61& $-14$     \\
FeI(1183)& 5565.71 &  0.89      &${\it -11}$ &     &           &      &           & 0.82& $-19$     \\
FeI(686) & 5572.84 &  0.74      &  $-9$      &     &           & 0.72 &${\it -2}$ &     &           \\
FeI(686) & 5586.76 &            &            & 0.63&  $-11$    & 0.72 &  $-2$     &     &           \\
FeI(686) & 5615.65 &            &            &     &           & 0.64 &${\it -3}$ & 0.62& $-22$     \\
FeI(1314)& 5633.95 &  0.89      & $-11$      & 0.86&${\it -11}$&      &       &     &           \\
SiI(10)  & 5645.61 &  0.94      & $-15$      &     &           &      &       &     &           \\
ScII(29) & 5669.03 &  0.87      &${\it -16}$ & 0.82&  $-12$    & 0.87 &${\it -1}$ &     &           \\
ScII(29) & 5684.19 &            &            &     &           &      &           & 0.76&${\it -22}$\\
NaI(6)   & 5688.21 &  0.76      &${\it -15}$ & 0.74&  $-11$    & 0.76 &${\it -4}$ &     &           \\
FeI(1107)& 5763.00 &            &            & 0.81&  $-14$    &      &       &     &           \\
SiI(17)  & 5772.15 &  0.92      &${\it -16}$ &     &           &      &       &     &           \\
SiI(9)   & 5793.07 &            &            &     &           & 0.91 &${\it -3}$ &     &           \\
CI(18)   & 5800.59 &  0.94      & $-15$      &     &           &      &           &     &           \\
FeI(982) & 5809.22 &  0.92      & $-16$      &     &           &      &       &     &           \\
BaII(2)  & 5853.68 &  0.79      &  $-9$      &     &           &      &       &     &           \\
NaI(1)   & 5889.95 &  3.3       & $-22$      & 4.9 &  $-21$    & 4.8  & $-21$     & 6.0 & $-21$     \\
         &         &  1.7       & $-18$      & 0.2 &  $-17$    & 0.2  & $-17$     & 0.2 & $-18$     \\
NaI(1)   & 5895.92 &  3.0       & $-21$      & 4.2 &  $-21$    & 4.2  & $-21$     & 5.4 & $-21$     \\
         &         &  1.8       & $-18$      & 0.3 &  $-17$    & 0.4  & $-17$     & 0.3 & $-18$     \\
SiI(16)  & 5948.54 &  0.86      & $-11$      &     &           &      &       &     &           \\
FeII(46) & 5991.37 &  0.86      &${\it -16}$ &     &           &      &       &     &           \\
CI       & 6014.85 &  0.95      &${\it -17}$ &     &           &      &       &     &           \\
NiI(27)  & 6016.64 &            &            &     &           & 0.90 &  $-4$     &     &           \\
FeI(1187)& 6024.06 &            &            &     &           & 0.86 &  $-3$     & 0.82& $-24$     \\
FeI(1018)& 6027.05 &            &            & 0.86&${\it -14}$& 0.90 &${\it -1}$ &     &           \\
FeI(1259)& 6056.01 &  0.89      &${\it -16}$ & 0.84&${\it -13}$&      &       &     &           \\
FeI(207) & 6065.49 &            &            & 0.73&  $-13$    & 0.83 &  $-1$     &     &           \\
CaI(3)   & 6122.22 &  0.81      & $-16$      & 0.72&  $-12$    & 0.77 &  $-2$     & 0.75& $-21$     \\
BaII(2)  & 6141.72 &  0.70      &  $-7$      & 0.52&   $-8$    & 0.61 &  $-3$     & 0.62& $-14$     \\
CaI(3)   & 6162.17 &            &            &     &           & 0.76 &  $-2$     &     &           \\
FeI(62)  & 6173.34 &            &            &     &           & 0.91 &${\it -2}$ &     &           \\
FeI(62)  & 6219.29 &  0.85      & $-16$      &     &           & 0.84 &${\it -3}$ &     &           \\
FeI(207) & 6230.73 &  0.83      & $-20$      &     &           &      &       &     &           \\
FeI(169) & 6252.56 &            &            &     &           & 0.85 &${\it -3}$ & 0.82& $-24$     \\
FeI(111) & 6254.26 &            &            &     &           & 0.82 &${\it -3}$ & 0.81& $-26$     \\
FeI(169) & 6256.37 &            &            &     &           & 0.86 &  $-2$     &     &           \\
FeI(62)  & 6265.14 &  0.88      & $-22$      & 0.85&  $-15$    & 0.87 &${\it -1}$ & 0.83&${\it -23}$\\
FeI(342) & 6270.23 &            &            &     &           &      &           & 0.83&${\it -27}$\\
FeI(169) & 6344.15 &            &            &     &           & 0.93 &${\it -4}$ &     &           \\
SiII(2)  & 6347.10 &  0.80      & $-15$      & 0.72&  $-11$    & 0.72 &  $-6$     & 0.75& $-21$     \\
FeII(40) & 6369.46 &            &            &     &           & 0.85 &  $-3$     &     &           \\
SiII(2)  & 6371.36 &  0.86      &${\it -23}$ & 0.78&${\it -13}$& 0.80 &  $-6$     &     &           \\
FeI(168) & 6393.61 &  0.87      & $-18$      & 0.76&  $-11$    &      &       &     &           \\
FeI(816) & 6400.01 &            &            &     &           &      &           & 0.71& $-25$     \\
FeI(816) & 6411.66 &  0.87      & $-23$      & 0.79&  $-12$    &      &       &     &           \\
FeII(74) & 6416.92 &  0.86      & $-17$      & 0.75&${\it -15}$& 0.80 &  $-1$     &     &           \\
FeI(111) & 6421.36 &  0.84      & $-20$      & 0.73&${\it -11}$&      &       &     &           \\
FeI(62)  & 6430.85 &            &            &     &           & 0.82 &${\it -3}$ &     &           \\
FeII(40) & 6432.68 &  0.84      &${\it -20}$ & 0.76&${\it -11}$& 0.79 &  $-2$     & 0.78& $-20$     \\
CaI(18)  & 6439.08 &  0.75      & $-20$      & 0.73&${\it -13}$& 0.77 &  $-2$     & 0.72& $-21$     \\
CaI(19)  & 6449.82 &  0.84      & $-15$      & 0.84&  $-12$    &      &           & 0.82& $-20$     \\
FeII(74) & 6456.38 &  0.74      & $-15$      & 0.67&  $-13$    & 0.70 &  $-3$     & 0.67& $-21$     \\
CaI(18)  & 6471.66 &  0.90      & $-20$      & 0.85&  $-18$    &      &       &     &           \\
H$\alpha$& 6562.81 &  0.75      &$-165$      & 0.76&  $-180$   & 0.81 &${\it -95}$& 0.75& $-150$    \\
         &         &  0.73      & $-23$      & 0.72&  $-13$    & 0.74 &${\it -11}$& 0.81& $-26$     \\
FeI(1253)& 6569.22 &            &            & 0.83&${\it -12}$& 0.87 &${\it -3}$ &     &           \\
CI(22)   & 6587.61 &  0.87      &${\it -28}$ &     &           &      &           &     &           \\
TiII(91) & 6606.95 &            &            &     &           & 0.93 &  $-2$     &     &           \\
FeI(206) & 6609.12 &  0.92      &${\it -19}$ &     &           &      &       &     &           \\
CI       & 6655.51 &  0.95      &${\it -14}$ &     &           &      &       &     &           \\
FeI(268) & 6677.99 &            &            & 0.77&  $-12$    &      &       &     &           \\
CaI(32)  & 6717.69 &  0.90      &${\it -17}$ & 0.88&  $-12$    &      &       &     &           \\
CI       & 7115.19 &  0.86      &${\it -26}$ &     &           &      &       &     &           \\
KI(1)    & 7664.87 &  1.6       & $-22$      &     &           &      &       &     &           \\
KI(1)    & 7698.97 &  1.6       & $-25$      &     &           &      &       &     &           \\
\end{supertabular}
\label{lines}
\end{center}

\begin{table}[ht]
\caption{Elemental abundances $\epsilon(X)$ in the atmosphere of QY\,Sge.
        Here $\sigma$ is the dispersion of abundances and n, the number
	of lines that were used in the computations. We adopt the elemental
	abundances $\epsilon(X)_\odot$ in the solar atmosphere from
	[\cite{Grev}]}
\medskip
\begin{tabular}{lrrcrrr}
\hline
X &$\epsilon(X)_\odot$&$\epsilon(X)$&$\sigma$&n&$[X/H]_\odot$ &$[X/Fe]$\\
\hline
C{\sc i}     & 8.55  &   8.94 & 0.09 &  9 & +0.39  & +0.25 \\
N{\sc i}     & 7.97  &   8.38 &      &  1 & +0.41  & +0.27 \\
Na{\sc i}    & 6.33  &   7.22 & 0.09 &  3 & +0.89  & +0.75 \\
Mg{\sc i}    & 7.58  &   7.74 & 0.03 &  2 & +0.14  & +0.00 \\
Si{\sc i}    & 7.55  &   7.71 & 0.23 & 21 & +0.16  & +0.02 \\
S{\sc i}     & 7.21  &   7.62 & 0.09 &  4 & +0.41  & +0.27 \\
Ca{\sc i}    & 6.36  &   6.43 & 0.12 &  8 & +0.07  &$-0.07$ \\
Sc{\sc ii}   & 3.17  &   3.16 & 0.22 &  5 &$-0.01$ &$-0.15$ \\
Ti{\sc i}    & 5.02  &   4.84 & 0.10 &  3 &$-0.18$ &$-0.32$ \\
Ti{\sc ii}   &       &   4.60 & 0.09 &  4 &$-0.42$ &$-0.56$ \\
V{\sc i}     & 4.00  &   4.20 & 0.05 &  2 & +0.20  & +0.06 \\
V{\sc ii}    &       &   4.16 & 0.04 &  3 & +0.16  & +0.02 \\
Cr{\sc i}    & 5.67  &   5.95 & 0.24 &  5 & +0.28  & +0.14 \\
Cr{\sc ii}   &       &   6.05 & 0.19 &  9 & +0.38  & +0.24 \\
Mn{\sc i}    & 5.39  &   5.74 & 0.05 &  3 & +0.35  & +0.21 \\
Fe{\sc i}    & 7.50  &   7.64 & 0.18 & 66 & +0.14  & --     \\
Fe{\sc ii}   &       &   7.64 & 0.26 &  8 & +0.14  & --    \\
Co{\sc i}    & 4.92  &   5.10 & 0.06 &  2 & +0.18  & +0.04 \\
Ni{\sc i}    & 6.25  &   6.49 & 0.19 & 12 & +0.24  & +0.10 \\
Cu{\sc i}    & 4.21  &   4.38 &      &  1 & +0.17  & +0.03 \\
Zn{\sc i}    & 4.60  &   4.64 & 0.12 &  2 & +0.04  &$-0.10$ \\
Y{\sc ii}    & 2.24  &   1.56 & 0.03 &  3 &$-0.68$ &$-0.82$ \\
Ba{\sc ii}   & 2.13  &   2.60 &      &  1 & +0.47  & +0.33 \\
Nd{\sc ii}   & 1.50  &   1.46 & 0.28 &  3 &$-0.04$ &$-0.18$ \\
Eu{\sc ii}   & 0.51  &   0.91 & 0.04 &  2 & +0.40  & +0.26 \\
\hline
\end{tabular}
\label{chem}
\end{table}

\begin{table}[hbtp]
\caption{Errors of the computed abundances of various chemical elements
        in the atmosphere of  QY\,Sge $\Delta\,log\,\epsilon(X)$, due
	to the uncertainty of the determination of the atmospheric
	parameters of the star}
\medskip
\begin{tabular}{l rrr}
\hline
&\multicolumn{3}{c}{$\Delta\,log\,\epsilon(X)$}\\
\cline{2-4}
X & $\Delta\,T_{eff}$   & $\Delta\,lg\,g$  & $\Delta\,\xi_t$  \\
  & $-$250\,K           & $-0.2$           &  $-0.5$\,km/s     \\
\hline
C{\sc i}   & $+$0.10 & $-$0.15 & $+$0.02 \\
N{\sc i}   & $+$0.18 & $-$0.15 & $+$0.03 \\
Na{\sc i}  & $-$0.12 & $-$0.09 & $+$0.13 \\
Mg{\sc i}  & $-$0.13 & $-$0.11 & $+$0.10 \\
Si{\sc i}  & $-$0.11 & $-$0.09 & $+$0.03 \\
S{\sc i}   & $+$0.05 & $-$0.12 & $+$0.04 \\
Ca{\sc i}  & $-$0.16 & $-$0.09 & $+$0.03 \\
Sc{\sc ii} & $-$0.08 & $-$0.14 & $+$0.12 \\
Ti{\sc i}  & $-$0.20 & $-$0.08 & $+$0.01 \\
Ti{\sc ii} & $-$0.07 & $-$0.14 & $+$0.08 \\
V{\sc i}   & $-$0.22 & $-$0.08 & $+$0.01 \\
V{\sc ii}  & $-$0.05 & $-$0.13 & $+$0.01 \\
Cr{\sc i}  & $-$0.20 & $-$0.10 & $+$0.06 \\
Cr{\sc ii} & $+$0.01 & $-$0.15 & $+$0.11 \\
Mn{\sc i}  & $-$0.18 & $-$0.09 & $+$0.08 \\
Fe{\sc i}  & $-$0.18 & $-$0.09 & $+$0.07 \\
Fe{\sc ii} &    0.00 & $-$0.14 & $+$0.09 \\
Co{\sc i}  &$-$0.22  & $-$0.09 & $+$0.01 \\
Ni{\sc i}  & $-$0.20 & $-$0.09 & $+$0.03 \\
Cu{\sc i}  & $-$0.25 & $-$0.10 & $+$0.06 \\
Zn{\sc i}  & $-$0.19 & $-$0.13 & $+$0.11 \\
Y{\sc ii}  & $-$0.09 & $-$0.13 & $+$0.02 \\
Ba{\sc ii} & $-$0.17 & $-$0.11 & $+$0.26 \\
Nd{\sc ii} & $-$0.14 & $-$0.13 & $+$0.01 \\
Eu{\sc ii} & $-$0.10 & $-$0.11 & $+$0.02 \\
\hline
\end{tabular}
\label{error}
\end{table}


\end{document}